\documentclass{aastex63} 
\usepackage{amsmath}
\usepackage{amssymb}
\usepackage[normalem]{ulem}
\usepackage{appendix}
\usepackage{graphicx}

\begin{document}
\shorttitle{GRO J2058+42 outburst in 2019 with AstroSat}
\shortauthors{Mukerjee et al.}
\title{AstroSat observations of GRO J2058+42 during the 2019 outburst}
\author[0000-0003-4437-8796]{Kallol Mukerjee}
\affiliation{Tata Institute of Fundamental Research, Homi Bhabha Road, Mumbai 400005, India}
\email{kmukerji@tifr.res.in}
\author[0000-0001-7549-9684]{H. M. Antia}
\affiliation{Tata Institute of Fundamental Research, Homi Bhabha Road, Mumbai 400005, India}
\author[0000-0002-9418-4001]{Tilak Katoch}
\affiliation{Tata Institute of Fundamental Research, Homi Bhabha Road, Mumbai 400005, India}

\begin{abstract}
We present  results from  AstroSat observation of the recent  outburst of  
GRO~J2058+42, an X-ray pulsar in a Be-binary system. The  source was 
observed on April 10, 2019 by LAXPC and SXT instruments on AstroSat during 
its declining phase of the latest giant outburst.  Light curves showed  
a strong  pulsation  of the pulsar with a period of $194.2201 \pm 0.0016$ s, and a spin-up
rate of $(1.65\pm0.06)\times10^{-11}$ Hz s$^{-1}$.
Intermittent flaring  was detected in light curves between 3--80 keV energy band with increase in 
intensity by up to 1.8 times its average intensity.  Pulse profiles 
obtained between 3--80 keV energy band  of the pulsar showed strong 
dependence on energy.  A broad  peak was observed in the 
power density spectrum of the source  consistently during  AstroSat observations  
with its peak oscillation  frequency of 0.090 Hz  along with its higher 
harmonics, which may be due to  quasi-periodic oscillations, a commonly 
observed phenomenon in transient X-ray pulsars, during their 
outburst. AstroSat observation  also detected  cyclotron absorption features in its
spectrum corresponding to (9.7--14.4) keV, (19.3--23.8) keV and (37.8--43.1) keV. 
The pulse phase resolved spectroscopy  of the source 
showed phase dependent variation in its energy and relative strength of these features.  
The spectrum was well  fitted with an absorbed  black-body,  a Fermi Dirac cut-off model 
and alternatively with an  absorbed CompTT model. Both these models were combined with a Fe-line  
and three Gaussian absorption lines to account for observed  cyclotron resonance scattering features
in the spectrum.
\end{abstract}

\section{Introduction}

Many  Be-binary systems  were observed earlier during outbursts which 
offered interesting results  \citep{bildsten97, reig11}.  
Bright X-ray outbursts are observed in Be-binaries most likely during
periastron  passage of its neutron star through a circumstellar disk of its companion.  
Depending on  amount of matter released  from its companion and  the geometry of the binary system, 
a rare  as well as regular outbursts are observed during its each binary orbit \citep{okazaki01, okazaki02, okazaki16}. 
The pulse characteristics of some of these were studied in detail during their 
outburst activities  such as EXO 2030+375 \citep{parmar89}, Cepheus X-4 \citep{mukerjee00}, 
XTE J1946+274 \citep{paul01}. 
Studies on pulse characteristics offer  information on the pulsar geometry 
and underlying mechanism for its emitted pulse profile.  The shape of the emitted  pulse  
depends on  modes of accretion inflows, source luminosity, geometry of
accretion column and configuration of its  magnetic field with respect to an observer's line of sight \citep{parmar89}. 
Therefore, such studies offer understanding of  its pulsar system and disk-magnetospheric interaction during 
the process of mass accretion, which affects its emitted radiation.       
Quasi-periodic oscillations (QPO) were detected from many  Be-binaries earlier, 
such as A0535+262 \citep{finger96, finger98}, EXO2030+375 \citep{angelini89}, 
4U0115+63 \citep{soong89, heindl99, dugair13}, V0032+53 \citep{qu05}. Studies of 
QPOs offer rich information about accretion torque onto the neutron star,  
thermodynamic properties of the inner accretion disk and electrodynamics of 
disk-magnetospheric interaction of the neutron star. Details on sources with observed
QPOs, their frequencies and other features along with pulsar spin frequencies etc.\  
are  given in  tabular form by \citet{devasia11, ghosh98, mukerjee01}.
Some of these transient Be-binary  pulsars such as A0535+26 (50 mHz, 9.7 mHz), 1A 1118-61 
(92 mHz, 2.5 mHz), XTE J1858+034 (110 mHz, 4.53 mHz), EXO 2030+375 (200 mHz, 24 mHz),
SWIFT J1626.6-5156 (1000 mHz, 65 mHz), XTE J0111.2-7317 (1270 mHz, 32 mHz) as well as
a persistent Be-binary, X-per (54 mHz, 1.2 mHz) and an OB-type binary, 
4U 1907+09 (69 mHz, 2.27 mHz) showed higher QPO frequency than their respective 
spin-frequency as  mentioned in order inside parenthesis \citep{devasia11}.
These cover  an interestingly wide range of QPO frequencies between 50--1270 mHz
for these pulsars.  Studies of cyclotron absorption features, if present 
in the spectrum, enable us to determine the strength of the surface magnetic field  
of the neutron star and offers an insight into the line producing region,
structure  of accretion column  and its geometry \citep{staubert19}. The cyclotron absorption features thus  
provides an important diagnostic probe for detailed studies of the neutron star binaries 
since  its discovery in the spectrum of Her X-1 \citep{truemper78}.   
There are several reports on the detection of cyclotron absorption features in the 
spectrum of many Be-binaries starting at a lower energy from $\sim10$ keV \citep{decesar13, jun12} 
to  a higher energy at $\sim100$ keV \citep{barbera01}. Detailed compilation of such
sources and  studies are reported by \citet{staubert19} and \citet{maitra16}. 
It is observed  from detailed studies that some sources show a wide variation in its 
cyclotron-line energy with respect to its pulse-phase, source luminosity  
and with time, such as Vela X-1, Cen X-3 and Her X-1 \citep{staubert19}. 
These interesting properties  help in understanding the nature of these sources and also offer 
an insight into their underlying physical properties governing such changes.

The high mass X-ray binary GRO J2058+42 is  a transient  X-ray pulsar
which was first discovered by BATSE on board CGRO during its  giant
outburst in  September--October, 1995.  The outburst lasted for 
about 46 days  which peaked  at 300 mCrab intensity \citep{wilson98}.  
The spin period of its neutron star decreased from 198~s to 196~s during 
its 46 days outburst \citep{wilson98}. This outburst was subsequently followed by 5 more
bursts of lower intensity of about 15 mCrab  and of lesser duration of  15 days  
observed at  an interval of  about 110 days.  Additional shorter 
outbursts with peak intensity of about 8 mCrab were detected by 
BATSE, halfway between the first four outbursts of  15-20 mCrab (20-50 keV) 
intensity.  During  early 1998, two outbursts 
of lower intensity were observed with PCA and HEXTE 
on-board RXTE \citep{wilson05}.  There  were, however, no reports of any giant 
outburst from the source. Thus, GRO~J2058+42 showed a very rare giant 
outburst activity, only twice so far, since its discovery. 
The first was detected by BATSE \citep{wilson98, wilson05} and 
the most recent one was detected by Swift-BAT \citep{barthelmy19, kennea19}
and Fermi-GBM \citep{malac19} in March 2019.  GRO J2058+42 was suggested earlier to be a 
high mass X-ray binary due to its observed properties.  It was subsequently  
confirmed  having  a Be-star companion through optical observations 
\citep{reig05}. A very limited details are known so far about this source, 
due to its rare intense outbursts.

  AstroSat  observed the source on April 10, 2019 during the declining phase of the  
second giant outburst, for studies of some of the above  described properties of  
a typical Be-binary pulsar.  In this work, we present results from our studies on spectral and 
timing properties of GRO J2058+42 using data from AstroSat observation.
Incidentally, there were no reports on detection of any cyclotron absorption features
from the  earlier outbursts of GRO~J2058+42. We, therefore  in particular,  wish to check for possible 
detection of cyclotron absorption features in the spectrum and presence of QPOs. 
After this manuscript was submitted, \citet{mol19} have reported detection 
of cyclotron absorption feature using NuSTAR observation of the source during the same 
outburst.  The rest of the paper is organized as follows: Section 2 describes the observations
and data analysis including software tools used, Section 3 describes the results 
and its implications are discussed in Section 4. Finally, Section 5 gives the 
conclusions from this study.
%
\begin{figure}
\epsscale{0.8}
\plotone{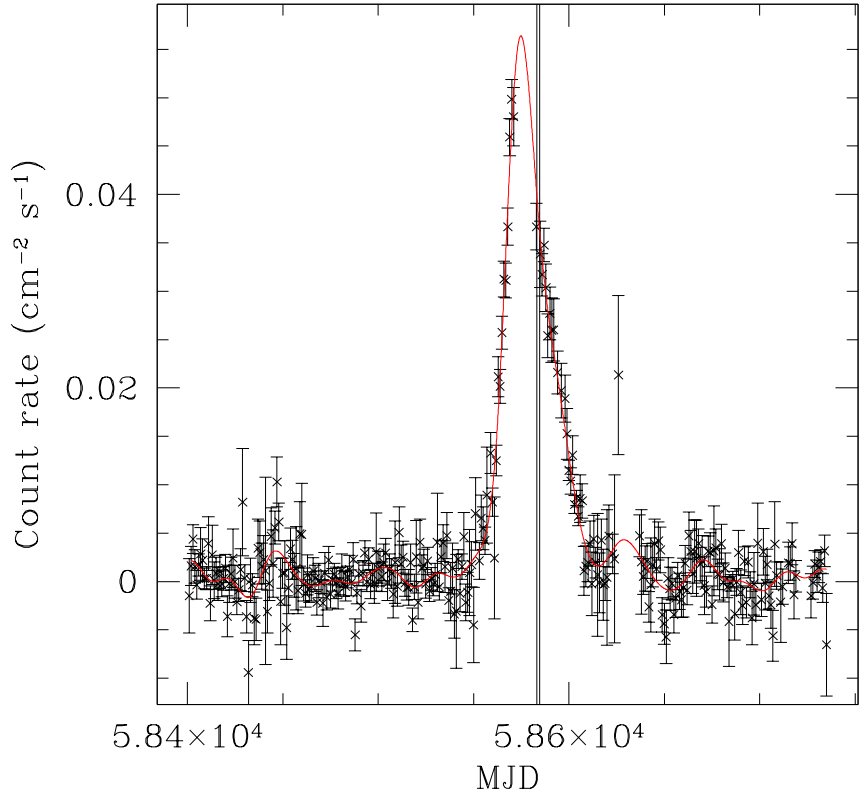}
\caption{Intensity variation in the 15--50 keV energy band as observed by Swift-BAT 
during source outburst is shown by black points while the continuous  line 
shows a fit to the light curve with cubic B-splines using 32 uniformly distributed knots. 
The time covered by AstroSat Observation is marked by two vertical lines.}
\label{lcfit}
\end{figure}
%
%
\section{Observations and Data Analysis }

GRO J2058+42  had its rare and long outburst during  March--May, 2019 as 
its latest event, as seen in the Swift-BAT intensity curve shown in Figure~\ref{lcfit}.
The light curve was downloaded from Swift-BAT light curves    
Archive\footnote{https://swift.gsfc.nasa.gov/results/transients/index.html} and  relevant details 
about these curves are described by \citet{krimm13}. 
As there  was a data gap close to the peak of the outburst,  the  
peak of the outburst was approximately determined by fitting a cubic B-spline with 32
uniformly distributed knots to the data while data points with large errors were 
dropped.  It enabled us to determine  peak intensity and its corresponding time 
which  was found to be at around MJD 58574.8  with its peak intensity  reaching 
to about 256 mCrab (Swift-BAT 0.0564 cts cm$^{-2}$ s$^{-1}$) in the 15--50 keV 
energy band. When we
compared this giant outburst with the earlier outburst, it turns out that,   
the main flare of this  giant outburst lasted for about a similar duration 
of 46 days \citep{wilson98} and  was followed by a 
relatively weaker and a much narrower secondary flare as seen in Figure~\ref{lcfit}. 
There is also a small burst about 110 days before the main burst as seen in the swift-BAT
light curve (Figure~\ref{lcfit}). CGRO--BATSE observed 
the earlier giant outburst with a peak intensity of about 300 mCrab 
\citep{wilson98, wilson05}, comparable to  that of its latest 
outburst which was also followed by a secondary flare observed after
about 52 days. 

AstroSat  observed the source  on April 10, 2019 during the declining phase 
of the outburst as marked in the Figure~\ref{lcfit} from MJD 58583.11 to MJD 58584.67.  
It was about  8-days after the peak of the outburst, when source intensity had
decreased to about 170 mCrab, which is about 66\% of its peak intensity as obtained
above.  The LAXPC was used as the primary instrument for this observation with a 
total effective
exposure of 57~ks.  Details of LAXPC instruments on board AstroSat and its operation modes 
are described in detail by \citet{antia17}. During the observation only one 
LAXPC detector, i.e., LAXPC20 was working properly.  LAXPCs were operated in its 
event-analysis mode.  LAXPC10 was operating at a very low gain and it 
barely detected pulsation in the source. LAXPC20 detected an average source 
count-rate of $225.6 \pm 0.1$ cts s$^{-1}$ in the 6--70 keV energy band. 
The SXT was operated in its photon counting mode during the whole observation, 
with  an effective exposure of about 21~ks and detected 
an average  count-rate of $3.78\pm0.01$ cts s$^{-1}$ in the 0.8--7 keV energy band.  
The SXT instrument along with its operational modes are described  in detail by  
\citet{singh16}.  Thus useful data  from  the AstroSat SXT and LAXPC20 were used 
for timing and  spectroscopic analysis, covering a total 
energy band of 0.8--80 keV. The AstroSat observation was made under a Target of 
Opportunity (TOO) proposal with the observation ID of 20190410\_T03\_098T01\_9000002836.  
AstroSat data from the proposal are available from the AstroSat 
Data Archive\footnote{https://astrobrowse.issdc.gov.in/astro\_archive/archive/Home.jsp}.
 
LAXPC data were analyzed using LAXPC pipeline software version 3.0. The
LAXPC
software can be downloaded from the LAXPC site\footnote{http::www.tifr.res.in/\~{}astrosat\_laxpc/LaxpcSoft.html}.
The software takes 
level-1 data files as input and
also utilizes calibration and background files to generate level-2 data products, like light
curve and spectrum.
The background files are available with the 
software, which also gives a suitable recommendation for selection of a  
background and associated response files which  may be used for data  product
generation and analysis.  The background observation during April 19, 2019 was used 
to estimate the contribution from the source.  Events were extracted from all 
main anodes from all the layers of LAXPC20 for this analysis. The pipeline
software was executed with  default values.  However, selection of appropriate
energy  channels were made for deriving light curves for a particular energy band.
The pipeline software also corrects for any shift in the gain based on calibration data
and generates  corrected spectrum and background subtracted light curves of the source
within its set energy band. 

The SXT data analysis pipeline software version AS1SXTLevel2-1.4b was used for 
a set of default parameters in this analysis. Spectral response file and 
auxiliary response files  of the SXT including background files are offered 
along with data analysis tools  and can be downloaded from the 
SXT
payload operation center site\footnote{http::www.tifr.res.in/\~{}astrosat\_sxt/index.html}.
Updated CALDB files are directly 
linked to the pipeline processing software.  All  photon events  were extracted  
by including a circular area covering  a region of interest  of 6 arc minutes 
radius with respect to the source center.  All grades between 0--12 were 
considered for selection of photon events.  The pipeline  analysis software  
takes  level-1 data  as its input along with other relevant files for the analysis 
of events in steps to finally produce clean and calibrated event lists.  
The final data products such as spectrum, light curve and image are produced  by 
applying various default filters and appropriate screening of the  data.  
From calibrated clean  events, one can produce spectrum and light curves 
in different energy bands covering 0.8--7.0 keV. The  `Xselect' tool 
version 2.4c  was used for screening the data. The solar system barycentric corrections  
were applied to the time series  for both SXT and LAXPC data  to correct 
for  arrival time delays of events prior to its detailed timing analysis, 
using a tool `as1bary'\footnote{http://astrosat-ssc.iucaa.in/?q=data\_and\_analysis},
developed by the AstroSat science support cell.

Heasoft version 6.14 was used for analysis of this data.  The standard software 
XRONOS 5.22 was used for timing  analysis; lcurve for  generation of combined 
light curves,   efsearch for  determination of 
spin-period of the pulsar, powspec for computing power density spectrum etc..  
Standard spectral analysis  tool XSPEC version 12.8.1 was used for fitting  
spectral data with combination of appropriate models as described below. 
Results  obtained from these analysis are presented in the next section.

\section{Results} 
\subsection{X-ray light curve and folded pulse profiles}
 
\begin{figure}
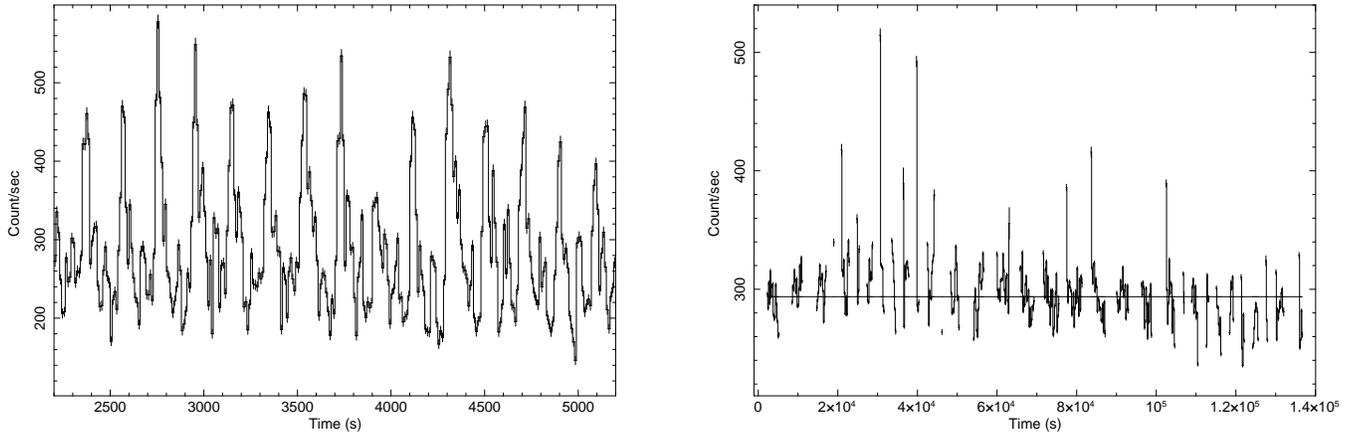

\centerline{\includegraphics[scale=0.35,angle=-90]{lc_10s.ps}
\qquad\includegraphics[scale=0.35,angle=-90]{lc_all_3-80keV.ps}}
\caption{Light curves derived from LAXPC20 in the 3--80 keV energy band are shown, as a segment with  10 s binning
(left panel) and also for the entire AstroSat observation with a binning of spin-period of the  pulsar (right panel). 
The reference time $t_0$ is MJD 58583.10868148 for both these figures. The horizontal line 
on the right panel shows the average source count rate of 294 counts s$^{-1}$.}
\label{lc2}
\end{figure}

Extracted light curve of GRO J2058+42 showed regular and strong pulsations 
along  with variation in  its intensity. A light curve segment of 
LAXPC20 with 10 s binning is shown in the left panel of Figure~\ref{lc2}.
Source intensity  variation  between the 3--80 keV energy band for  the complete observation duration of 
LAXPC20 is shown  in the right panel of Figure~\ref{lc2}, obtained by 
binning the data with its established spin-period.  A straight-line over 
the  binned light curve shows its average intensity.  Intensity variations  
were clearly seen for the  complete  
duration of AstroSat observation.  Observed  source count rate varied 
from 230 cts s$^{-1}$ to about 520 cts s$^{-1}$ with an average 
count rate of  294 cts s$^{-1}$. Intermittent flaring and dips were 
also  seen in the light curve which is a typical intrinsic property of a Be-binary. 
During the flares,  intensity increased by up to 1.8 times the average intensity.

The pulsar spin period  was derived  after applying corrections to pulse arrival 
time delays with respect to solar system barycenter.
The average pulsar period  was initially  derived with epoch folding and verified using
Lomb-Scargle periodogram,  on full AstroSat observation data. Spin-period of the 
pulsar was found to be $194.180\pm0.001$ s with 1-$\sigma$ confidence limit, during AstroSat observation. 
Total duration of the observation time was then divided into 5 intervals 
to estimate spin-up rate of the pulsar during  AstroSat observation. 
The pulsar spin-period was determined for each data 
interval separately along with its  estimated  error. These periods were 
plotted with respect to  mid-value of  respective interval in MJD.  
An average spin-up rate of $(1.71\pm0.14)\times10^{-11}$ 
Hz s$^{-1}$ was derived by  fitting a straight line to these 5 data points,
along with its error estimated with 90\% confidence limit.  Then  starting with these initially measured values,  
the spin period at the beginning of observation  at $t=t_0$ and its time derivative were accurately determined
by correcting the phase using
\begin{equation}
	\phi(t)=\phi_0+\nu_0(t-t_0)+\dot\nu{(t-t_0)^2\over 2},
	\label{eq:phase}
\end{equation}
where $\phi$ is the phase (in range 0--1 over the period), $\phi_0$ is the phase at initial time, $t_0$,
$\nu_0$ is the spin frequency at initial time and $\dot\nu$ is its time derivative.
The fit was performed by fitting a periodic signal with 20 harmonics of the basic
period using  Eq.~\ref{eq:fit},
\begin{equation}
	c(t)=c_0+\sum_{k=1}^N\Big(a_k\sin(2\pi k\phi)+b_k\cos(2\pi k\phi)\Big),
	\label{eq:fit}
\end{equation}
where $c(t)$ is the observed count rate, $N$ is the number of harmonics included in the fit
and $c_0,a_k,b_k$ are the coefficients fitted, apart from $\nu_0$ and $\dot\nu$.
The light curve was obtained with a time-bin of 1 s. The best fit values for $\nu_0$ and
$\dot\nu$  were determined by varying both parameters to minimize $\chi^2$  deviation of the
light curve from the model (Eq.~\ref{eq:fit}).  Various values of $N$ were tried and
$N=20$ was found to be adequate to account for all pulsation signal. In principle,
it is possible to use the F-test to decide the required value of $N$, which gave a value of
$N=14$ with a probability threshold of 0.05, but some components after $k=15$ were also significant. Beyond $N=20$, the
next few components were not found to be significant. That is why this value was adopted. To find the
errors in the fitted values of $\nu$ and $\dot\nu$, we performed a Monte Carlo simulation
by perturbing $c(t)$ and repeating the process for 4000 different realization of noise
to find the distribution of parameter values and using this distribution, we found the 90\%
confidence limits for the parameters. The same program was used to generate the pulse-profiles
as well as the GTI intervals for different phase intervals using the fitted values of parameters.
The GTI values were then used to calculate the spectra for different phase intervals for
phase resolved spectroscopy. The same program was also used to generate the light curve
after subtracting the pulse profile as defined by Eq.~\ref{eq:fit} to filter out the
contribution of coherent pulsation in the resulting power density spectrum.

The fitted value of the period at $t=t_0$  corresponding to MJD 58583.10868148 was found to be $194.2201\pm0.0016$ s
and $\dot\nu=(1.65\pm0.06)\times10^{-11}$ Hz s$^{-1}$, where the error bars denote the 90\%
confidence limits.
The orbit of the binary system is not known and the orbital motion can 
also contribute to the period variation during the observation. However, the Fermi-GBM 
observations during this outburst\footnote{https://gammaray.msfc.nasa.gov/gbm/science/pulsars/lightcurves/groj2058.html}
suggests that the period variation was restricted to the duration of the giant outburst, and
the change in ${\dot \nu}$ during the outburst showed a good correlation with intensity, with a
correlation coefficient of about 0.93. This implies that the variation in period 
is largely due to accretion during the outburst.
%
%
\begin{figure}
\centerline{\includegraphics[scale=0.75]{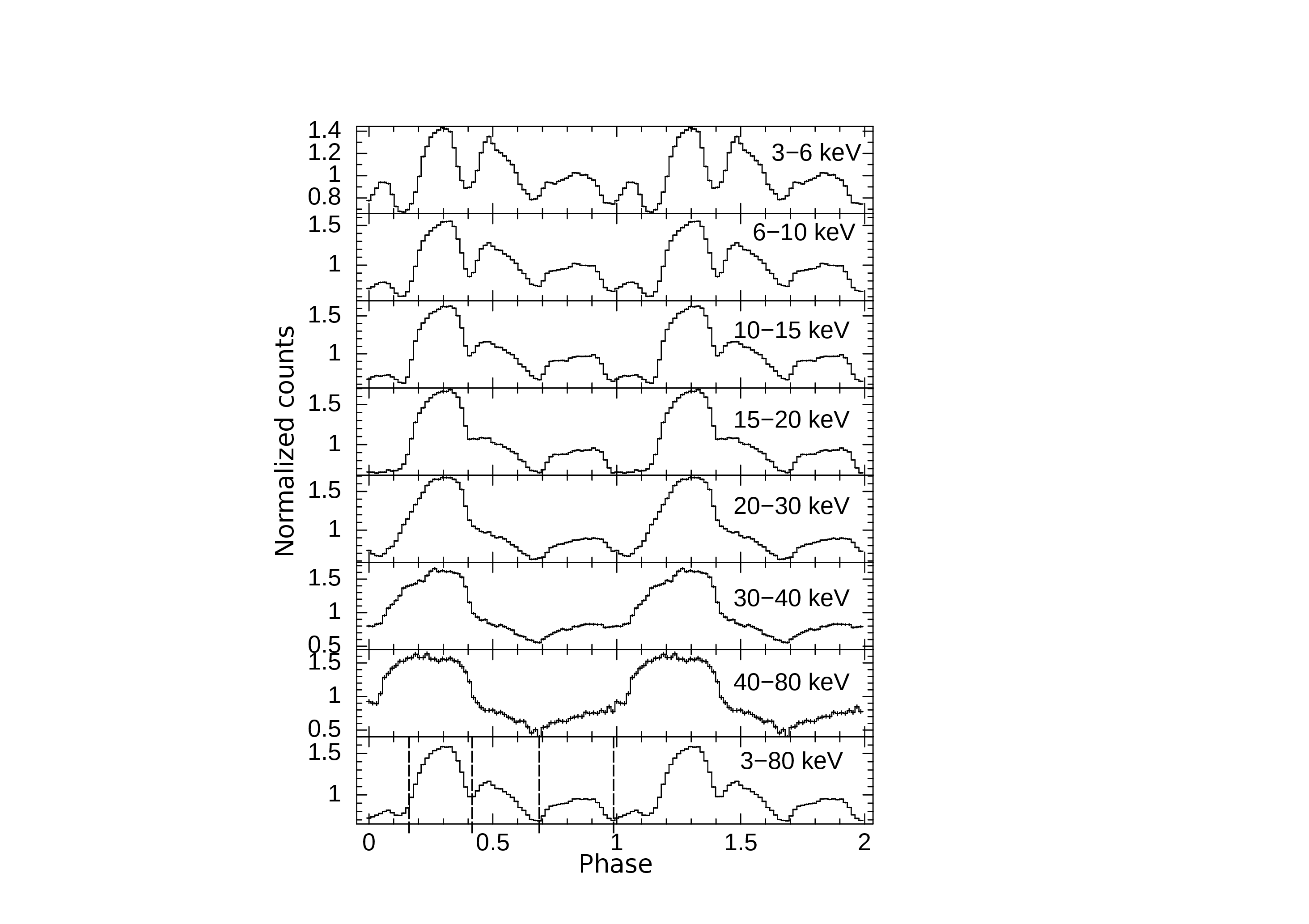}} 
\caption{Folded pulse profiles  derived from LAXPC20 at different energy bands 
covering 3--80 keV. The vertical lines in the lowest panel mark the four divisions of
pulse period used in phase resolved studies.}
\label{pulse}
\end{figure}
%
%
\begin{figure}
\centerline{\includegraphics[scale=0.5,angle=-90]{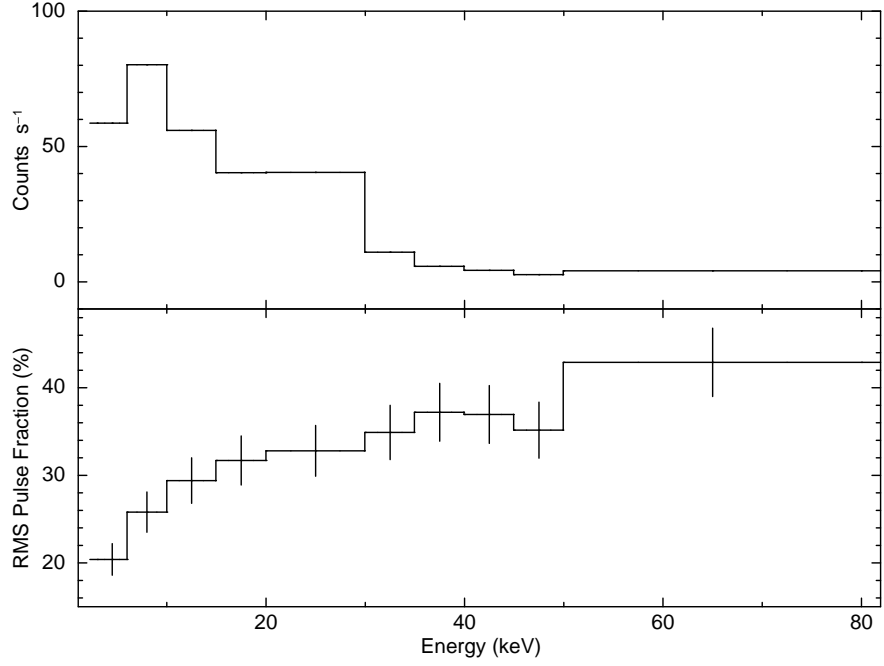}}
\caption{Count rate and RMS pulse  fraction derived from LAXPC20 at different energy bands 
covering 3--80 keV.}
\label{rmspf}
\end{figure}
%

Light curves were extracted  corresponding to  different energy bands and respective 
pulse profiles were derived by folding  corresponding  light curves in 64 bins  with its 
derived spin-period of the neutron star at the beginning of the epoch and its derivative. Two cycles of all such pulse-profiles  
are shown in Figure~\ref{pulse} for clarity.  Folded pulse-profiles showed 
variation  with respect to  different energy bands.  
The average pulse-profile  covering full 3--80 keV energy band is  also 
shown at the bottom panel of the Figure~\ref{pulse}, for reference and 
relative comparison.  We noticed that pulse profiles at lower energies showed  
pronounced  and multiple-pulses which gradually  merged into a double-pulse 
and then in a single asymmetric broad-pulse at higher energies.  
The RMS pulse fraction  measurements and  source intensity with respect 
to energy is shown in Figure~\ref{rmspf}. 
RMS pulse fraction showed energy dependent variation,  which showed increase 
from  about $20\%$ to $30\%$ between  3--20 keV while at higher energies it is nearly constant.
Thus pulse profiles showed  remarkable 
variability in shape and its pulse-fraction with respect to  considered 
energy bands between 3--80 keV.  

Pulse-profiles derived from recent AstroSat observations can be compared 
with earlier RXTE observation at much lower source intensity of about 10 mCrab.  
There is a general similarity in shape but structures are 
seen much more clearly than those reported from 
November 28, 1996 RXTE observation between 2--60 keV during its 
earlier lower intensity outburst \citep{wilson98}.  
For the source intensity of about 17 mCrab, the RMS pulse fractions were measured at 
different energy bands between 2--20 keV from the RXTE PCA observations of February 4, 1998. 
These were estimated  as   $11.7 \pm 2.6 \%$ (2--4 keV), $19.1 \pm 4.1\%$ (4--9 keV) and 
$27.2 \pm 5.9 \%$ (9--20 keV). This can be compared 
with the recent observation by  AstroSat LAXPC20, which gives
$15.6 \pm 3.3 \%$ (3--4 keV), $22.0 \pm 4.7 \%$ (4--9 keV) and $28.6 
\pm 6.1 \%$ (9--20 keV) respectively.  
The RXTE/PCA and AstroSat/LAXPC20 measurements show that these are comparable
within respective error limits, although source intensity was an order of magnitude
higher during AstroSat observation. Thus, the pulsar does not show drastic
change in shape of pulse profiles and pulse-fractions  
with change in source luminosity in this range.

\begin{figure}
\centerline{\includegraphics[height=8.0 cm]{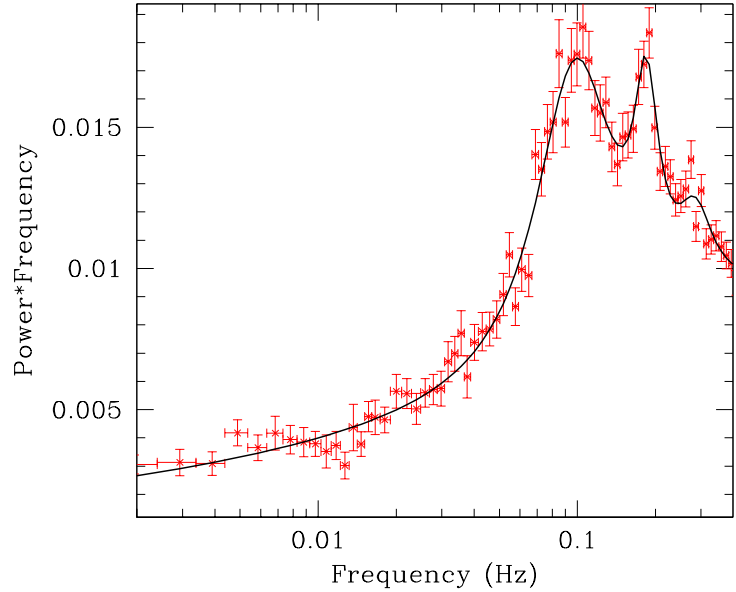}} 
\caption{Power density spectrum  derived from LAXPC20 data between 3--30 keV 
covering the observation duration is shown. 
A prominent QPO  feature is observed at 0.09 Hz along with its 2 harmonics 
when signal from coherent  spin-frequency of neutron star was 
removed from the time series.}
\label{qpo}
\end{figure}
\subsection{Power density spectrum and detection of a QPO feature}

The source light curve along with its background between  
3--30 keV energy band with a bin time of 1.0 second was used to derive  
power density spectrum. The XRONOS tool `powspec' was used for this purpose. The
normalization parameter value of $-2$ was selected  so that power density spectra 
were normalized such that their integral gave the squared RMS fractional variability.
Therefore, the power  was expressed in the units of $\mathrm{(RMS)}^2\, \mathrm{Hz}^{-1}$ and
the expected white noise level was subtracted, to obtain the RMS fractional variability 
of the time series. For such a normalized power density spectrum where a QPO profile was 
modeled by  a Lorentzian, the strength of a  QPO signal was described by its fractional 
root-mean-squared (RMS) amplitude, which is proportional to its integrated power which contributes 
to its over all power density spectrum and often expressed in percent \citep{wang14}.
The integrated power can be  computed from the area under the Lorentzian profile.
The amplitude of the Lorentzian  multiplied by its FWHM and $\pi$/2 would determine the area
under the profile in units of $\mathrm{RMS}^2$, hence its square-root would give the RMS amplitude
of the QPO signal.  The power density spectrum was therefore, derived considering above normalization and
a total of  1024 bins per interval and  total a 73 intervals in a frame 
were considered for the purpose of  obtaining the power density spectrum. However, 
out of a total of 73 intervals, 16 intervals with less than $50\%$ data,
were rejected for a default window selection.  The data gaps were padded with 
zeros as its  default option.  A geometric re-binning of 1.04 was applied for generation of 
power density spectrum  to improve on statistical fluctuation at higher frequencies.

\begin{deluxetable}{ccccc}
\tablecolumns{2}
\tablecaption{Power density spectrum model parameters}
\tablehead{\colhead{Parameter} &\colhead{Value}
&\colhead{Amplitude} &\colhead{F-TEST} &\colhead{Detection}\\
&    & (RMS~$\%$) & FAP &Significance ($\sigma$) }
\startdata
Power Index & $-0.78\pm0.08$ &  & &    \\
Normalization (K) (at 1 Hz)   &$0.010\pm 0.001$  &  & & \\\\
QPO-$f_{0}$ Centroid LC (Hz)  & $0.090\pm 0.003$ & $ 12.1 \pm 1.0$ & $2.7 \times 10^{-11}$  & 6.7 \\
Sigma LW (Hz) & $0.081\pm0.012$ & & & \\
Normalization LN  & $0.116 \pm 0.008$ & &\\\\
QPO-$f_{1}$ Centroid LC (Hz) &$0.183 \pm 0.004$ & $5.2 \pm 0.9$ & $1.4 \times 10^{-7}$ & 5.3\\ 
Sigma LW (Hz) & $0.048 \pm 0.015$ & & &  \\
Normalization LN & $0.036\pm 0.006$ & & &\\\\
QPO-$f_{2}$ Centroid LC (Hz) & $0.280 \pm 0.017$ & $ 4.2 \pm 1.5$ & $2.2 \times 10^{-4}$ & 3.7 \\
Sigma LW (Hz) & $0.11 \pm 0.07$ & & & \\
Normalization LN & $0.010 \pm 0.003$ & & &\\
\hline
$\chi^2_{{red}}$ (dof)   & 1.08 (62) & & &   \\
\hline
\enddata
\tablecomments{Errors with 90\% confidence range for each parameter. FAP-False alarm probability.}
\end{deluxetable}
%

The average power density spectrum  derived as above  showed spin frequency
along with its several  harmonics. It also showed a broad peak around 0.09 Hz and its 2 harmonics, 
which were identified as a QPO.  The QPO signal was  found  to be better in the 3--30 keV 
energy band, compared to a higher energy band, hence  power spectra  are presented for 
this energy band. 
For confirming the  presence of QPO, 
we removed the contribution of  coherent pulsation signal by subtracting
the fitted  pulse profile with 20 harmonics (Eq.~\ref{eq:fit}) from the time series data as described in section~3.1. 
Power density spectrum  was then derived from this time series. 
The power density spectrum showed a clear detection of a QPO feature at $0.090 \pm 0.003$ Hz 
along with its 2 harmonics.   As the pulsation signal was removed 
from the time series data, power density spectrum did not show any coherent pulsations as 
was seen earlier. 
However, the  continuum defined by the power law-index were found to be consistent
within the 90\% error limit, when
coherent pulsation signal was present (power-index $= -0.89 \pm{0.09}$) and  when pulsation signal was removed,
(power-index $= -0.78 \pm{0.07}$). 
The observed  QPO and its 2 higher harmonics defined by
Lorentzian function were also found consistent within their parameter uncertainty.  
The power density spectrum was 
modeled using a  power-law in combination with 3 Lorentzian functions
to model QPO and its two  higher harmonics. The model parameter values and QPO-frequency and its corresponding
2 harmonics as derived from the fitted model are listed in the Table-1. This is to mention here that since 
many intervals of data were averaged for generating the power density spectrum, therefore, $\chi^2$ test for the fitted model does
not introduce significant biases as it is well known that in this case the power density spectrum has distribution close 
to the normal distribution \citep{barret12}.
The model was fitted and  parameter values 
were established from the power density spectrum when its y-axis was  normalized  and expressed in the units of
$\mathrm{(RMS)}^2\, \mathrm{Hz}^{-1}$ and  frequency 
along the x-axis. However, the power density spectrum is shown in Figure~\ref{qpo} with y-axis multiplied by 
corresponding frequency to show prominence of the QPO and its 2 harmonics. 
The Lorentzian function $L(f)$ used in the  model is defined as:
\begin{equation}
L(f) = \frac{\mathrm{LN}}{1 + \left(\frac{2 (f-\mathrm{LC})}{\mathrm{LW}}\right)^2} \;,
\end{equation}
where  parameters LN, LC and LW represent  its peak amplitude, centroid frequency and its 
full width at half maximum, respectively. 

An alternate model,  broken-power-law  along with 3 Lorentzian functions was also fitted to 
the same power density spectrum for comparison.  The reduced ${\chi^2}$ value of 1.12 was found for 
the 60  degrees of freedom (dof), as compared with power-law model along with 3 Lorentzian 
functions which yields  the reduced ${\chi^2}$ of 1.08 for 62 dof.  Thus, power-law model  
resulted in a better fit, compared to broken power-law model. 
A break frequency of $0.31 \pm 0.01$ Hz was obtained from  broken-power law model when fitted
to the power density spectrum.  The break  frequency is much higher
than the observed QPO frequency of 0.09 Hz and hence it cannot mimic the QPO.

Thus, RMS amplitudes  determined from  the averaged  power density spectrum, by fitting
a power-law  and 3 Lorentzian for QPO and its 2 harmonics are also given 
in Table-1.  The statistical significance of detected QPOs was established using the 
F-test. This was done  by determining the difference in $\chi^2$  with and without the 
inclusion of  a Lorentzian function, used for modeling detected QPO in the power density spectrum. 
Similar approach was followed for its other 2 harmonics, individually. 
The false alarm probability (FAP) and detection 
significance of QPO expressed in terms of $\sigma$ for an equivalent normal 
distribution is computed and listed in Table-1 for QPO and its 2 harmonics. It turns 
out that the QPO frequency and the two harmonics are significant. It is observed that apart from 
the first harmonic the other two peaks are rather broad. 

\begin{figure}
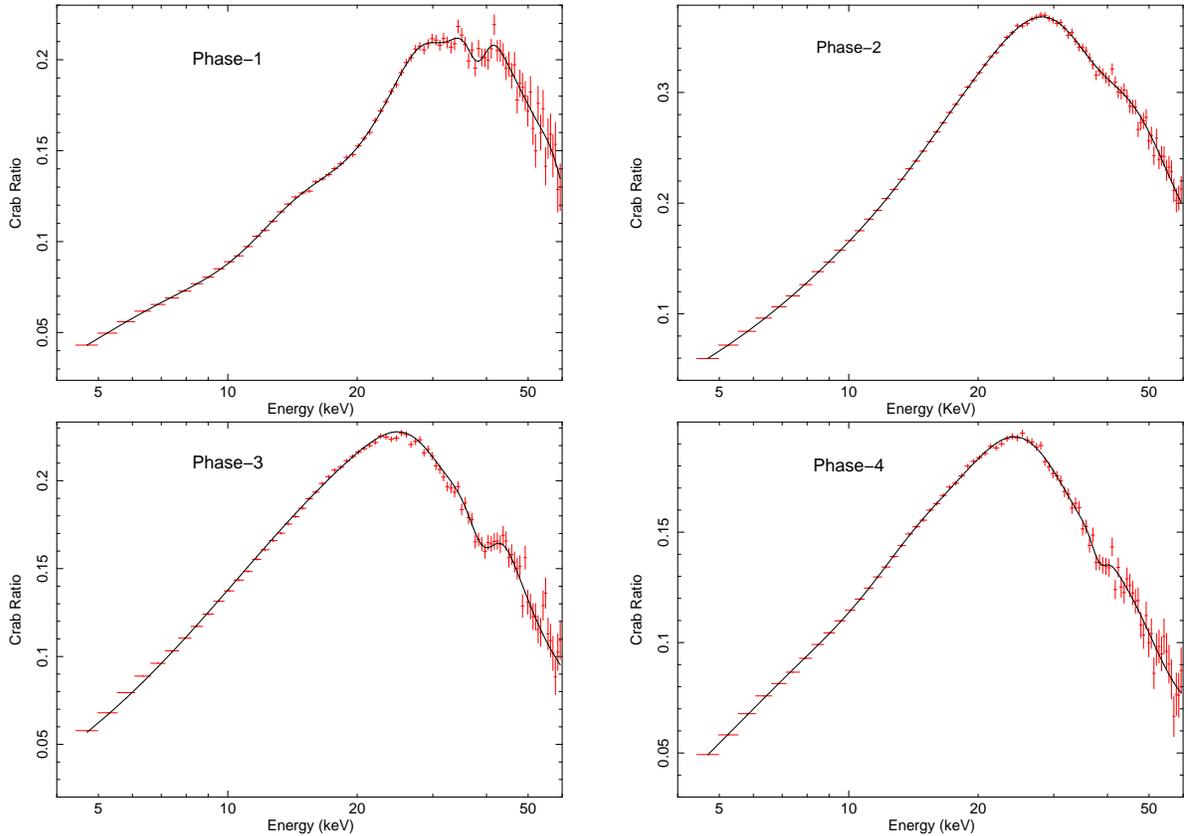

\centerline{\includegraphics[scale=0.36,angle=-90]{Crab_ratio_ph1_pl.ps}
\quad\includegraphics[scale=0.36,angle=-90]{Crab_ratio_ph2_pl.ps}}
\centerline{\includegraphics[scale=0.36,angle=-90]{Crab_ratio_ph3_pl.ps}
\quad\includegraphics[scale=0.36,angle=-90]{Crab_ratio_ph4_pl.ps}}
\caption{Ratio of spectral counts of the source at different energies  with respect to
Crab spectral counts is shown corresponding to 4 different pulse-phases of the pulsar (as
marked in Figure 3) derived from LAXPC20 spectral data between 4--60 keV energy band. 
The continuous line shows the fit with a combination model of a polynomial
of degree 3 in energy and Gaussians to define dips in the ratio.}
\label{crabr}
\end{figure}

\subsection{Spectrum and detection of cyclotron absorption features}

RXTE derived the first X-ray spectrum  of the source   
during the earlier outburst of 1996 covering energy band between 2.7--50 keV.  
The spectrum was modeled adequately well with absorbed thermal 
bremsstrahlung (Phabs*bremss). However, there was, no report
on any detection of cyclotron absorption feature from RXTE observations.
Therefore, to check for the presence of a cyclotron absorption feature in the spectrum we took the
ratio of counts in the spectrum of source to that in the Crab spectrum which
was observed by  LAXPC20 of AstroSat and the results for the phase-resolved spectra are shown in
Figure \ref{crabr}.
The overall pulse phase was  divided into 4 intervals 
covering structures of the folded pulse profile as shown by vertical lines
on 3--80 keV pulse profile in Figure~\ref{pulse}. The phase intervals considered are
0.00--0.18 (phase-1), 0.18--0.43 (phase-2), 0.43--0.70 (phase-3) and 0.70--1.00 (phase-4).
The ratio of spectral counts  plotted against energy with respect to Crab  was fitted with a polynomial 
of degree 3 along  with combination of Gaussians to define corresponding features 
in the ratio curve.   
The first phase shows two features around 10 keV and 20 keV
and an insignificant feature around 38 keV. The feature around 38 keV is more prominent during other
phases as well as in the phase averaged case, while the lower two features are
not significant during the last two phases. This is borne out by the results of
fitting the spectrum. The variation in properties of features with phase effectively rule out instrumental
effects or uncertainties in instrumental response as these effects would be independent
of phase.

%
%
\begin{deluxetable}{cccc}
\tablecolumns{4}
\tablecaption{Spectral parameters for phase averaged spectrum} 
\tablehead{\colhead{Parameter}& \colhead{Units}& \colhead{Model 1 (FDCUT)}& \colhead{Model 2 (CompTT)}}
\startdata
Phabs(nH) & 10$^{22}$ cm$^{-2}$ & $0.869^{+0.045}_{-0.046}$  & $0.650^{+0.048}_{-0.046}$ \\
bbody(kT) & keV & $0.83^{+0.04}_{-0.05}$ & - \\
bbody(Norm) & ph keV$^{-1}$ cm$^{-2}$ s$^{-1}$  &$0.0036^{+0.0008}_{-0.0007}$ & - \\
power-law(PI) &- & $1.04^{+0.03}_{-0.03}$ & - \\
$E_{cut}$ & keV & $34.6^{+1.2}_{-2.1}$ &  - \\
$E_{fold}$ & keV & $10.95^{+0.27}_{-0.43}$ & -  \\
power-law(Norm) & ph keV$^{-1}$ cm$^{-2}$ s$^{-1}$ & $0.10^{+0.01}_{-0.01}$ & - \\
 & at 1 keV & & \\
CompTT(Redshift)  & - & - & 0 (fixed) \\
CompTT(T0)  & keV & - & $0.51^{+0.02}_{-0.03}$ \\
CompTT(kT)  & keV & - & $8.22^{+0.10}_{-0.10}$ \\
CompTT(tau)  & - & - & $5.21^{+0.12}_{-0.12}$ \\
CompTT(approx) & -  & - &  0.2 (fixed) \\
CompTT(Norm)  & -  & - & $0.051^{+0.003}_{-0.002}$ \\
gauss(Line E) & keV & 6.5 (fixed) & 6.5 (fixed) \\ 
gauss(Sigma) & keV & 0.24 (fixed) & 0.24 (fixed) \\
gauss(Norm)  & $ph~cm^{-2}~s^{-1}$ & $1.8\times 10^{-4}$  & $2.3\times 10^{-4}$ \\
gabs(Line E1) & keV & $10.95^{+0.74}_{-0.81}$ & $10.81^{+0.48}_{-0.49}$ \\ 
gabs(Sigma) & keV & $3.74^{+0.72}_{-0.53}$  &  $3.20^{+0.39}_{-0.32}$  \\
gabs(Strength)  & -   & $1.13^{+0.29}_{-0.29}$  & $1.56^{+0.28}_{-0.29}$ \\
gabs(Line E2) & keV & $21.71^{+0.81}_{-0.83}$ & $20.84^{+0.54}_{-0.55}$ \\ 
gabs(Sigma) & keV & $3.61^{+1.28}_{-1.32}$  &  $3.69^{+0.47}_{-0.40}$  \\
gabs(Strength)  & -   & $0.90^{+0.48}_{-0.44}$  & $1.42^{+0.27}_{-0.29}$ \\
gabs(Line E3) & keV & $38.73^{+0.90}_{-0.81}$  & $38.38^{+1.14}_{-1.08}$\\
gabs(Sigma) & keV & $2.64^{+1.73}_{-1.33}$ & $2.56^{+1.12}_{-0.77}$ \\
gabs(Strength) & - & $1.02^{+0.55}_{-0.33}$ & $0.96^{+0.21}_{-0.20}$\\
Flux(3--80 keV) & erg cm$^{-2}$ s$^{-1}$  & $4.33\times 10^{-9}$ & $4.33\times 10^{-9}$ \\
\hline
$\chi^2$   & -         & 879.79 & 877.36  \\
$\chi^2_{{red}}$ (dof) &-  & 1.264(696) & 1.257(698)  \\
\hline
\multicolumn3c{Significance of cyclotron absorption features from F-TEST:}\\
\hline
E1 FAP & -  & $1.68 \times 10^{-2}$ & $1.79 \times 10^{-4}$ \\ 
 significance ($\sigma$) & - & 2.39  & 3.75  \\ 
E2 FAP & -  & $2.57 \times 10^{-4}$ & $1.72 \times 10^{-4}$ \\ 
significance ($\sigma$) & - & 3.66  & 3.76  \\ 
E3 FAP & - & $1.37 \times 10^{-7}$ & $1.55 \times 10^{-7}$ \\ 
significance ($\sigma$) & - & 5.27 & 5.25  \\ 
\hline
\multicolumn3c{Significance of cyclotron absorption features from NONE-ZERO LINE DEPTH: }\\
\hline
E1 FAP & -  & $1.19 \times 10^{-3}$ & $3.14 \times 10^{-6}$ \\ 
significance ($\sigma$) & - & 3.24  & 4.66  \\ 
E2 FAP & -  & $4.57 \times 10^{-6}$ & $5.93 \times 10^{-6}$ \\ 
significance ($\sigma$) & - & 4.49  & 4.36  \\ 
E3 FAP & - & $5.89 \times 10^{-13} $ & $2.27 \times 10^{-10}$ \\ 
significance ($\sigma$) & - & 7.20 & 6.34  \\ 
\enddata
\tablecomments{Errors with 90\% confidence range for each parameter. False alarm probability
 (FAP).} 
\end{deluxetable}
%
\begin{figure}
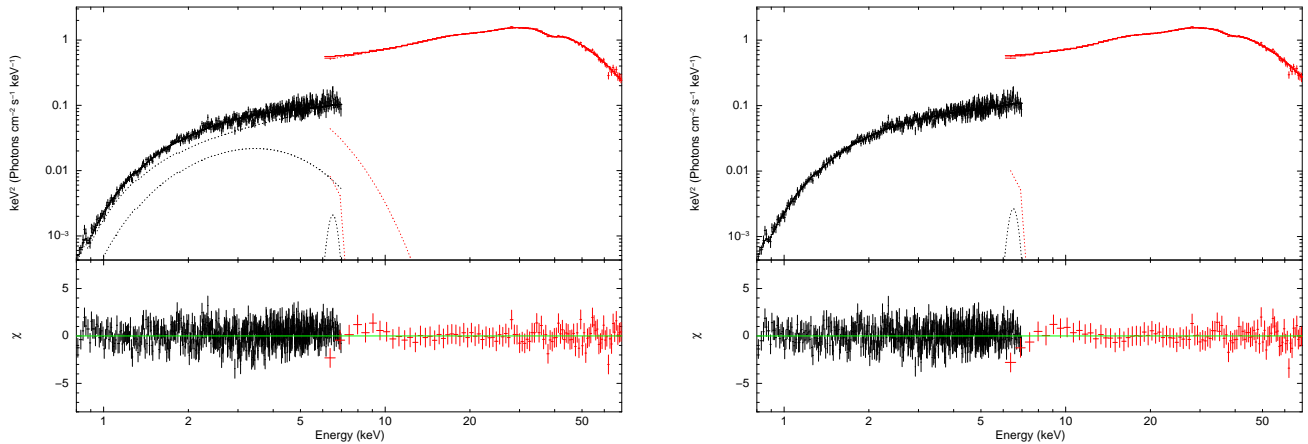

\centerline{\includegraphics[scale=0.34,angle=-90]{Spec_fdcut_av_feb.ps}
\quad\includegraphics[scale=0.34,angle=-90]{Spec_comptt_av_feb.ps}}
\caption{Spectrum  fitted with  combined model; an absorbed Fermi-Dirac cutoff model
with a black-body a Gaussian emission line and 3 Gaussian absorption lines (Model~1, left panel),  and 
an absorbed CompTT with a Gaussian emission line and 3 Gaussian absorption lines 
(Model~2, right panel) are shown.}   
\label{spec}
\end{figure}
%
%
\begin{figure}
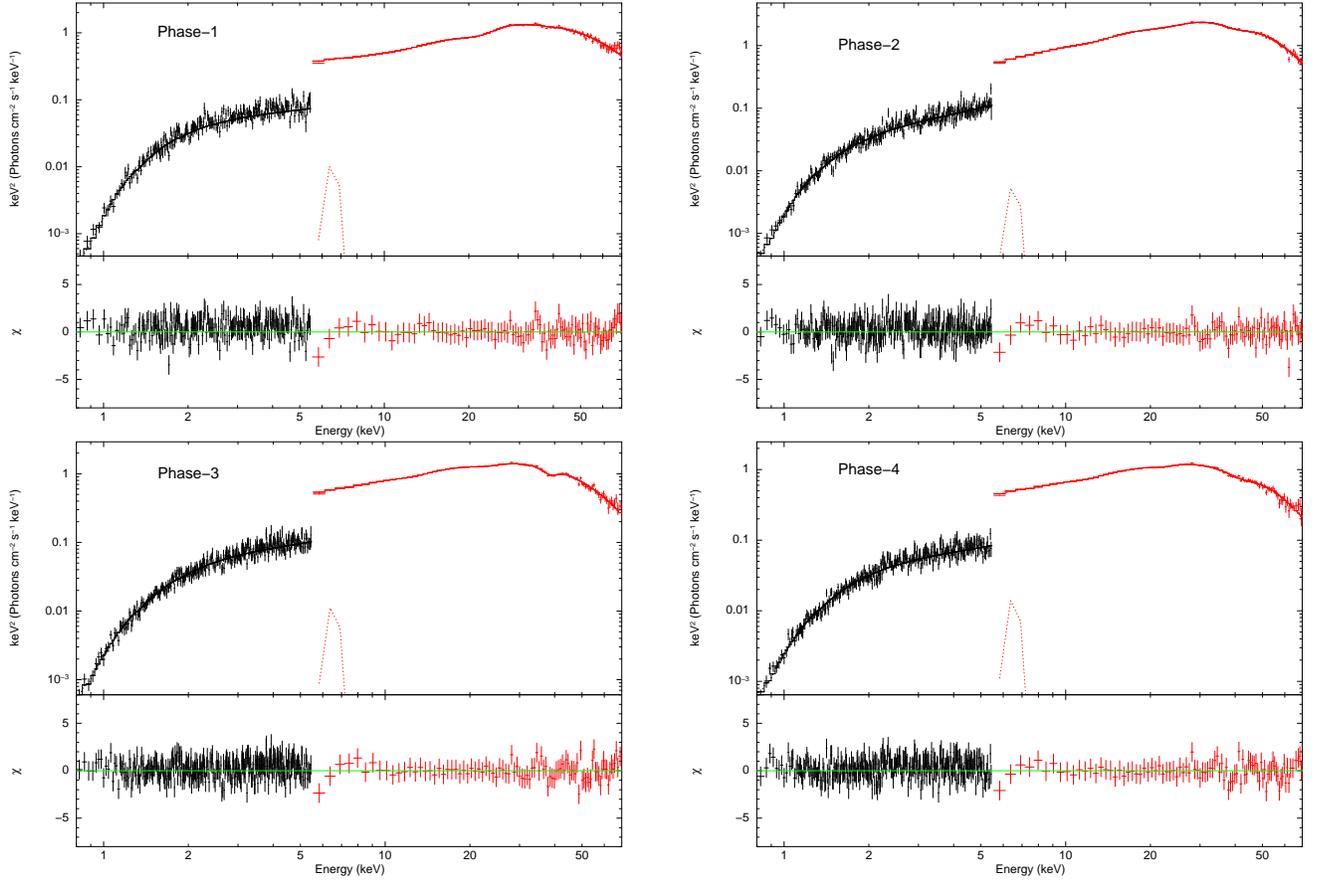

\centerline{\includegraphics[scale=0.34,angle=-90]{Spec_PH1_comptt_1.ps}
\quad\includegraphics[scale=0.34,angle=-90]{Spec_PH2_comptt.ps}}
\centerline{\includegraphics[scale=0.34,angle=-90]{Spec_PH3_comptt_m.ps}
\quad\includegraphics[scale=0.34,angle=-90]{Spec_PH4_comptt.ps}}
\caption{Phase resolved spectra  fitted with  combined model;  an absorbed  CompTT with 
a Gaussian emission line and
3 Gaussian absorptions lines.
The graph below shows the residue to the fit in  units of chi.}
\label{cspec}
\end{figure}
%
%
\begin{deluxetable}{cccccc}
\tablecolumns{6}
\tablecaption{Spectral parameters for phase resolved spectra} 
\tablehead{\colhead{Parameter}& \colhead{Units}& \colhead{Phase 1}& \colhead{Phase 2}& \colhead{Phase 3}& \colhead{Phase 4}}
\startdata
Phabs(nH) & 10$^{22}$ cm$^{-2}$ & $0.83^{+0.11}_{-0.11}$  & $0.81^{+0.19}_{-0.12}$  &$0.65^{+0.10}_{-0.08}$ & $0.64^{+0.09}_{-0.08}$ \\
CompTT(Redshift)  & - & 0 (fixed) & 0 (fixed) & 0 (fixed) & 0 (fixed) \\
CompTT(T0)  & keV & $0.41^{+0.05}_{-0.05}$  & $0.42^{+0.06}_{-0.11}$ & $0.53^{+0.04}_{-0.05}$ & $0.47^{+0.04}_{-0.04}$ \\
CompTT(kT)  & keV & $10.03^{+0.22}_{-0.22}$ & $8.56^{+0.10}_{-0.11}$ & $8.45^{+0.15}_{-0.26}$ & $8.36^{+0.15}_{-0.18}$ \\
CompTT(tau)  & - & $4.60^{+0.17}_{-0.16}$  & $6.17^{+0.16}_{-0.15}$ & $4.91^{+0.27}_{-0.13}$ & $4.99^{+0.11}_{-0.23}$ \\
CompTT(approx) & -  &  0.2 (fixed) &  0.2 (fixed)  & 0.2 (fixed)  & 0.2 (fixed) \\
CompTT(Norm)  & -  & $0.037^{+0.003}_{-0.003}$ & $0.055^{+0.006}_{-0.003}$ & $0.047^{+0.003}_{-0.002}$ & $0.042^{+0.007}_{-0.002}$ \\
gauss(Line E) & keV & 6.5 (fixed) & 6.5 (fixed)  & 6.5 (fixed) & 6.5 (fixed) \\ 
gauss(Sigma) & keV & 0.24 (fixed) & 0.24 (fixed)  & 0.24 (fixed) & 0.24 (fixed) \\
gauss(Norm)  & $ph~cm^{-2}~s^{-1}$   & $2.10\times 10^{-4}$  & $2.20\times 10^{-4}$ & $2.88\times 10^{-3}$  & $2.88\times 10^{-3}$  \\
gabs(Line E1) & keV & $10.26^{+0.51}_{-0.52}$ & $13.17^{+1.18}_{-0.99}$ & $12.44^{+1.55}_{-1.18}$ & $12.05^{+1.00}_{-0.89}$  \\
gabs(Sigma) & keV & $3.59^{+0.57}_{-0.48}$  &  $2.02^{+1.30}_{-1.01}$  & $1.49^{+1.72}_{-0.84}$  &  $1.28^{+1.26}_{-0.75}$  \\
gabs(Strength)  & -   & $2.14^{+0.58}_{-0.52}$  & $0.47^{+0.20}_{-0.26}$  & $0.20^{+0.26}_{-0.12}$  & $0.19^{+0.11}_{-0.13}$ \\
gabs(Line E2) & keV & $19.72^{+0.39}_{-0.40}$ & $22.20^{+0.70}_{-0.78}$  & $22.68^{+1.13}_{-1.15}$ & $22.11^{+1.27}_{-1.33}$  \\ 
gabs(Sigma) & keV & $3.66^{+0.58}_{-0.53}$  &  $3.68^{+0.61}_{-0.50}$  & $2.15^{+1.41}_{-0.89}$  &  $1.89^{+1.81}_{-1.21}$  \\
gabs(Strength)  & -   & $2.21^{+0.57}_{-0.44}$  & $1.21^{+0.60}_{-0.46}$  & $0.37^{+0.62}_{-0.17}$  & $0.27^{+0.17}_{-0.16}$ \\
gabs(Line E3) & keV & $38.38 (fixed)$  & $40.07^{+1.15}_{-1.11}$ & $38.54^{+0.70}_{-0.71}$  & $41.77^{+1.35}_{-1.90}$ \\
gabs(Sigma) & keV & $2.2 (fixed) $ & $4.60^{+0.87}_{-0.73}$  & $2.41^{+1.11}_{-1.08}$ & $5.33^{+1.45}_{-1.29}$ \\
gabs(Strength) & - & $0.059^{+0.21}_{-0.059}$ & $1.69^{+0.35}_{-0.36}$ & $1.15^{+0.24}_{-0.23}$ & $2.39^{+0.98}_{-0.75}$ \\
\hline
$\chi^2$   & -         & 663.48 & 528.95  & 526.29 & 625.61    \\
$\chi^2_{{red}}$ (dof) &-  & 1.211(548) & 0.969(546)  & 0.966(545)  & 1.148(545)  \\
\hline
$\chi^2_{{diff}}$(with 3 gabs removed)   & -         & 224.0 & 68.76  & 66.54 & 62.79    \\
FAP & - & $6.08 \times 10^{-30}$ & $5.42 \times 10^{-11}$  & $1.24 \times 10^{-10}$  & $3.67 \times 10^{-8}$ \\ 
significance ($\sigma$) & - & 11.37 & 6.55 & 6.43  & 5.51 \\ 
\hline
\multicolumn4l{Significance test of cyclotron line using F-TEST:}\\
\hline
E1 FAP & - & $9.67 \times 10^{-13}$ & $4.13 \times 10^{-4}$  & $9.29 \times 10^{-2}$  & $9.83 \times 10^{-3}$ \\ 
significance ($\sigma$) & - & 7.14 & 3.53 & 1.68  & 2.58 \\ 
E2 FAP & - & $2.22 \times 10^{-7}$ & $1.58 \times 10^{-6}$  & $2.51 \times 10^{-3}$  & $1.02 \times 10^{-1}$ \\ 
significance ($\sigma$) & - & 5.18 & 4.80 & 3.02  & 1.64 \\ 
E3 FAP & - & $9.9 \times 10^{-1}$ & $7.26 \times 10^{-13}$  & $1.56 \times 10^{-9}$  & $6.11 \times 10^{-9}$ \\ 
significance ($\sigma$) & - & 0.01 & 7.17  & 6.04  & 5.81  \\ 
\hline
\multicolumn4l{Significance test of cyclotron line using NON-ZERO LINE DEPTH:}\\
\hline
E1 FAP & - & $4.05 \times 10^{-17}$ & $1.34 \times 10^{-4}$  & $5.13 \times 10^{-2}$  & $3.81 \times 10^{-3}$ \\ 
significance ($\sigma$) & - & 8.41 & 3.82 & 1.95  & 2.89 \\ 
E2 FAP  & - & $1.76 \times 10^{-9} $ & $3.89 \times 10^{-7}$  & $1.24 \times 10^{-3}$  & $2.27 \times 10^{-2}$ \\ 
significance ($\sigma$) & - & 5.73 & 4.96 & 3.23  & 2.21 \\ 
E3 FAP & - & $9.26 \times 10^{-1}$ & $1.38 \times 10^{-13}$  & $5.70 \times 10^{-14}$  & $3.01 \times 10^{-11}$ \\ 
significance ($\sigma$) & - & 0.09 & 7.40  & 7.08  & 6.65  \\ 
\enddata
\tablecomments{Errors with 90\% confidence range for each parameter. False alarm probability (FAP)}.
\end{deluxetable}
%

\begin{figure}
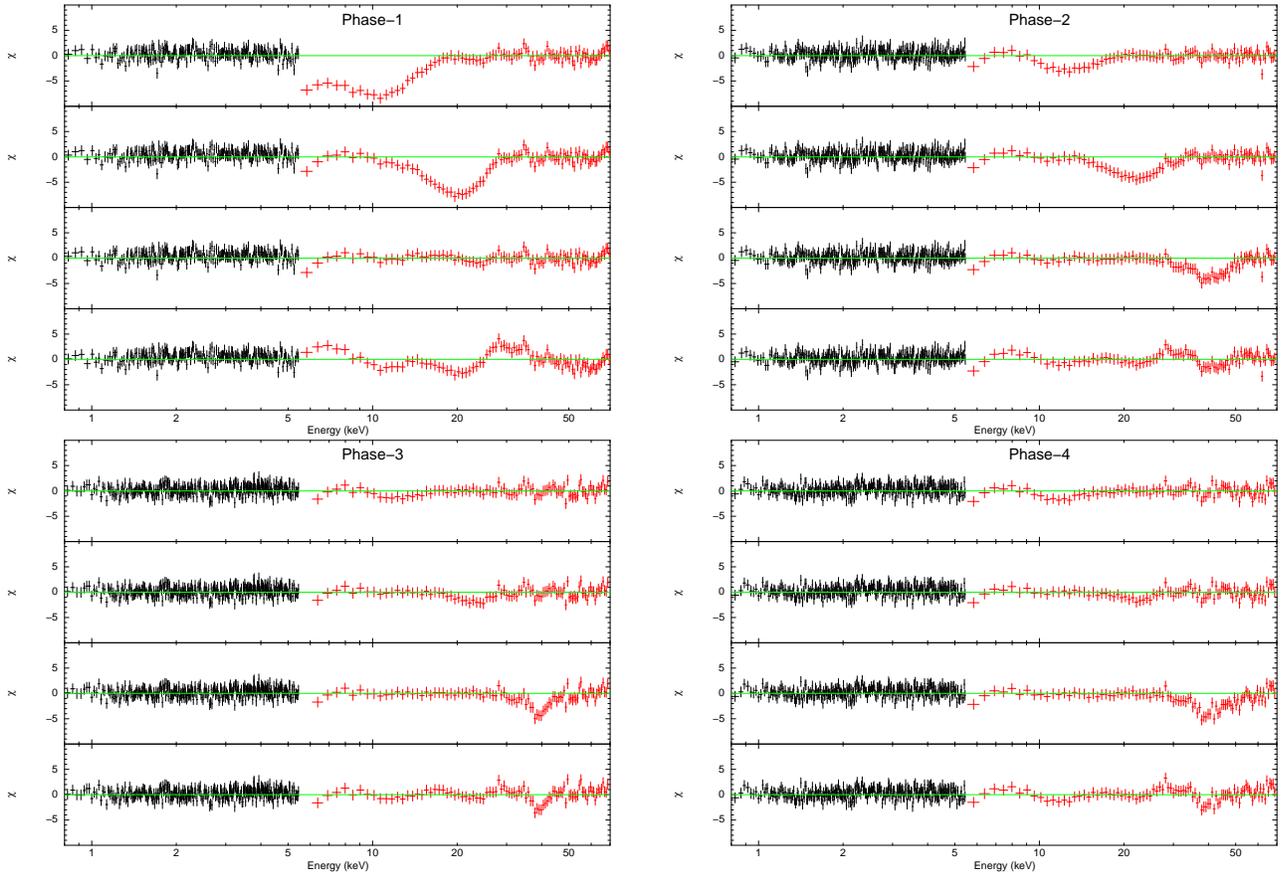

\centerline{\includegraphics[scale=0.34,angle=-90]{PH1_comb_delc_fitted.ps}
\quad\includegraphics[scale=0.34,angle=-90]{PH2_comb_delc_fitted_m.ps}}
\centerline{\includegraphics[scale=0.34,angle=-90]{PH3_comb_delc_fitted.ps}
\quad\includegraphics[scale=0.34,angle=-90]{PH4_comb_delc_fitted.ps}}
\caption{Phase resolved spectra residues showing relative intensities 
of 3 absorption lines (top 3 panels of each figure) when individual  optical depth is made zero and 
followed by fitted residue when all the three absorption features were removed from the model (bottom panel of each figure)
for all the 4 different phases. The combined spectrum was fitted with Model~2.}
\label{cspec1}
\end{figure}

We  then derived  X-ray spectrum covering a total energy band from 
0.8--70 keV  with AstroSat, combining  SXT and LAXPC20 spectral 
data.  The  combined spectrum was fitted with two different models. 
The first (Model--1) is an absorbed Fermi-Dirac cutoff model,
FDCUT~\citep{tanaka86} combined with a black body (bbody), 
a Gaussian for an iron emission line and  3 Gaussian absorption 
features (gabs)  which  were introduced to account for presence of  cyclotron 
absorption features in the spectrum  to improve the spectral 
fit.  We also tried an alternative model (Model--2) 
an absorbed CompTT model \citep{tit94} combined with an iron emission 
line and 3 absorption features as described above for comparison and  
measurement of the centroid energy of the cyclotron absorption features. 
The spectral parameters derived from these two models are tabulated 
in Table~2 for phase-averaged  spectrum, while Table~3 shows the results of fitting
the second model to phase-resolved 
spectra.  Spectra along with the fitted models are shown 
for phase-averaged and phase-resolved spectra in Figures~\ref{spec} and \ref{cspec},
respectively. 
The Figure~\ref{cspec1} shows phase dependent variations in 
spectral residues indicating relative intensities of 3 absorption features 
individually  when the optical depth of corresponding line energy is made zero. In addition,  
the residue of the  fitted model is also presented  when all the  3 gabs components,  associated with cyclotron  
resonance scattering features, were removed. This depicts phase dependent variations of relative intensities of
cyclotron absorption features for 4 different phases. The difference in $\chi^2$ values without the  3 gabs components
were established for all the 4 phases and given in the Table~3. The difference in $\chi^2$ values clearly showed
overall significance of presence of cyclotron absorption features in the spectrum.  A systematic error of $2.5\%$ was added  
to account for uncertainties in the response of the 
instruments.  Relative  difference in  the normalization between instruments  
was accounted  by  introducing a constant multiplicative factors for the two instruments and by 
fixing it for LAXPC20 at 1.0 during the fit. 
Thus the normalization factor  of SXT was found to be 0.31 when a circular field of view
of  a diameter of 40 arc minutes was  selected.  This is due to a fixed offset between the pointing axis of the two instruments. 
The region of interest of SXT, however,  was restricted to 12 arc-minutes  in diameter
to  accumulate events mainly from region where source photons dominate, to allow better
constrain in the lower energy  region of the overall spectrum.
With this restricted region of interest, the normalization was found to be 0.18 for SXT with respect to the LAXPC20 frozen at unity.
Both the spectral models offered a reasonably good fit to 
the spectral data and  measurements of respective energy of cyclotron 
absorption features are found to be consistent within their error limits.  
Considering the observed flux of $4.3\times 10^{-9}$ ergs cm$^{-2}$ 
s$^{-1}$ in the 3--80 keV energy band and using source distance of 
9 kpc \citep{reig05}, the X-ray luminosity is determined to be
$4.2 \times 10^{37}$ ergs s$^{-1}$ during AstroSat observation.

To estimate the statistical significance of the cyclotron absorption features which 
were detected, we  initially tried  first two approaches described by \citet{bhaler15} 
and then cross verified these using Monte Carlo technique as a re-verification  
for a few cases, as described below.  

\begin{figure}
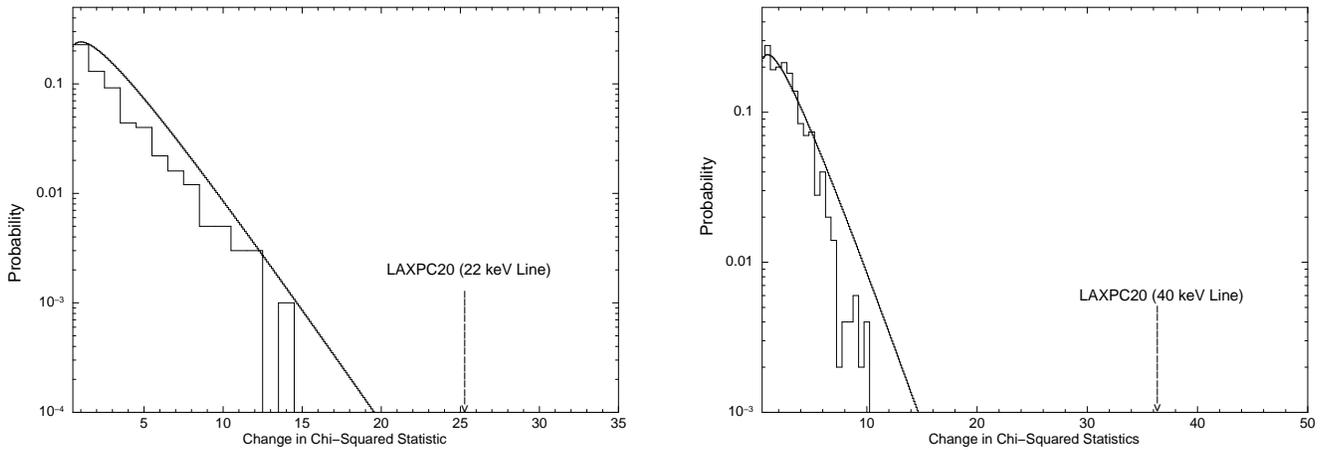

\centerline{\includegraphics[scale=0.34,angle=-90]{E2_sim.ps}
\quad\includegraphics[scale=0.34,angle=-90]{E3_sim.ps}}
\caption{Results of $\Delta\chi^2$ distribution derived from Monte Carlo simulations of spectra for LAXPC20 for testing 
significance of cyclotron absorption features corresponding to lines at $\sim 22$ keV (left panel) 
and $\sim 40$ keV (right panel) are shown for pulse phase-2.  The observed values of $\Delta\chi^2$ corresponding
to  above two lines are shown by  vertical arrows.} 
\label{simft}
\end{figure}

\begin{enumerate}
\item The first approach was to use the F-test, which uses the reduction in
$\chi^2$ value when the three parameters defining the cyclotron absorption feature  using 
gabs (multiplicative model) were included. Based on the improvement in $\chi^2$ value 
by adding the feature, one could estimate its significance and false alarm probability. 
However, this is to specifically mention that the F-test used here is additive and has 
certain limitations for multiplicative models such as gabs-model \citep{orlan12,prota02}.
This test was applied for an initial assessment and for rough estimates of the 
significance of 3 individual  additional components, defined by gabs-model. 
In all cases, the significance of each feature was calculated independently and we have 
calculated the probability of the signal being false and 
converted that to the significance in terms of equivalent value of $\sigma$ in the
normal distribution.  The estimated significance by this F-test for the absorption features in the spectrum  
are tabulated for phase averaged and  4 different pulse phases in Tables~2 and 3, 
respectively.

\item In the second approach, we kept the  centroid energy fixed for a cyclotron absorption 
feature and stepped through  a grid of values  of  line-depth and width in steps 
to study the change in $\chi^2$ as a function of these two parameters, using the XSPEC 
steppar command. The change in the minimum  $\chi^2$ value corresponding to zero line-depth
was noted.
Thus, using the minimum difference in $\chi^2$ required to get zero line-depth, we could calculate the 
false alarm probability and the  significance of the line. The results are  shown in Table~2 
for  phase averaged spectrum and in Table~3 for four different  pulse-phases of the pulsar. 
These estimates of significance of cyclotron absorption features are consistent with the estimates
using option (1).
 
\item We also assessed the significance of cyclotron absorption features independently 
using Monte Carlo simulation.  This technique has been used to establish 
significance of cyclotron absorption features in the spectrum for many sources 
\citep{bellm14,bhaler15,bodag16}.
We used LAXPC20 data fitted with the CompTT continuum model 
along with Fe-line  and 3 Gaussian absorption features which were introduced to model observed cyclotron 
absorption features. Only LAXPC20 data were considered to improve on the speed  
of simulation and convergence to the fit. We fixed the Hydrogen column density 
($N_{H}= 0.81\times 10^{22}$ cm$^{-2}$) as derived from fit to the combined SXT and LAXPC20 spectral data.  
We then used  XSPEC script `simftest'  for  simulating
1000 spectra. The  script does the fit with and without cyclotron absorption feature for each of
the simulated data.
From these simulations, one can  estimate the change 
in $\chi^2$  between a model  with and without the cyclotron absorption feature
and find the distribution of $\Delta\chi^2$ values. 
These simulations were done  to re-confirm the significance results for Phase-2 spectrum,  
corresponding to its $\sim$ 22 keV  and $\sim$ 40 keV cyclotron  absorption features, independently.  
The change in chi-squared 
($\Delta\chi^2$) distribution were plotted. Since the cyclotron absorption feature  is defined by three 
free parameters, we expect the   simulated $\Delta\chi^2$ distribution to follow a 
$\chi^2$ distribution with three degrees of freedom.  The observed distribution was found
to be consistent with the $\chi^2$ distribution as  can be  seen in the Figure~\ref{simft}.   
In these two cases maximum values of $\Delta\chi^2$ obtained in simulation was  14 (22 keV line) and 10 (40 keV line) 
which are much lower than the observed values of 25.5 (22 keV line) and 36.7 (40 keV line) in the 
LAXPC20 data.  By extrapolating the distribution we estimate that  a very large number of simulations  
would be required to get observed value of $\Delta\chi^2$ in one of them by chance. 
This could be $10^5$--$10^6$ simulations  required for  $\Delta\chi^2\approx 26$ and $10^{8}$--$10^{9}$  
for $\Delta\chi^2\approx 37$. Therefore, it is  not feasible to  perform these many simulations to 
achieve desired values.  The significance of individual features for LAXPC20 data  for phase-2, corresponding
to these values are  $4.38\sigma$ ($\mathrm{FAP}=1.2\times 10^{-5}$) for 22 keV line and $5.4\sigma$
($\mathrm{FAP}=5.32 \times 10^{-8}$) for 40 keV line.  These simulations therefore confirmed 
presence of  these features with high significance in LAXPC20 spectrum which detected  cyclotron  
absorption features from the source in its sensitive energy band.  
It  thus also  confirms  that the significance established from the first two methods as shown in the Table 3 
are consistent with the results  obtained by the  Monte Carlo simulation technique as shown here as a test case.
We have tried a few more cases and in all of these the significance is consistent with
those obtained using options (1) and (2).
      
\end{enumerate}

Our results for phase-1 are in agreement  with \citet{mol19}, whereby  the first two 
features  around 10 keV and 20 keV  were detected with high significance and no other 
higher energy features were detected.
The mean energy, width and depth of the features estimated by us is also consistent
with those by \citet{mol19}.
Additionally, we detected absorption feature around 38--40 keV 
for the  rest of the  3 phases with high significance.  However, we did not detect presence of any
cyclotron absorption feature around 30 keV as reported by \citet{mol19}.  This was verified
using phase-1 spectrum, where a third gabs model was introduced  with frozen  values of its energy  and width 
as reported by \citep{mol19} and its optical depth was allowed to vary. The  additional gabs-model fitted to the 
spectral data could not  detect any significant optical depth corresponding to  the energy $\sim$30 keV.  
The individual spectrum,  Figures~4 \&~5 
of \citet{mol19}, do not appear to show significant feature around 30 keV. 
The feature seen around 38--40 keV in LAXPC spectrum, could possibly be
either the third harmonic (if the primary absorption feature is around 12 keV) or the fourth harmonic
(if the primary feature is around 10 keV). In the latter case, we may have missed the third
harmonic as it could possibly  be very weak and hence not detectable. Even higher harmonics could 
be present, but due to lower sensitivity at high energies it is difficult to say anything 
definitely about them.
We  thus detected a cyclotron absorption feature around 38--40 keV in the other 3 phases, which 
is contrary to the results of  \citet{mol19}. This detection was possible  due 
to relatively  higher effective area of the LAXPC20 \citep{antia17} by more than an order of magnitude 
around 38--40 keV compared to FM-modules of the NuSTAR \citep{harison13,brejnholt12}.  
This is evident from their Figure 4 and 5, \citep{mol19}, where errors are relatively 
large around 38 keV and hence it is likely that 38 keV feature was not detected
by  the NuSTAR.  We detected 38 keV absorption  feature with high significance as 
reported in  Table-2,  even in the phase averaged spectrum. We also noticed that the 
spectral ratio with respect to Crab spectrum 
(Figure~\ref{crabr}) clearly showed depression  around 38--40 keV, confirming  presence of this 
absorption feature in the spectrum and the variation in its shape with phase. This  rules out the 
possibility of  occurrence of absorption feature due to inaccuracy
of instrument response or the modeled background.

\section{Discussion}
\subsection{Power Spectrum and QPO}

   Many accretion powered pulsars show  milli-Hertz quasi periodic 
oscillations (QPOs).  X-ray pulsars with high mass companions such  as 
X Per \citep{takeshima98}, 4U 1907+09 \citep{zand98, mukerjee01}, 
XTE J1858+034 \citep{paul98},  A 0535+26 \citep{finger96}, V 0332+53 
\citep{takeshima94} and X 0115+63 \citep{soong89} showed QPOs with  
their respective peak oscillation frequencies in the range 50--110 mHz.
QPOs in  X-ray pulsars offer an important diagnostic tool to probe; 
accretion flows in these binaries,  conditions  in the inner region 
of its accretion disk,  properties of accretion torques exerted on 
its neutron stars and  disk-magnetospheric coupling and their 
interaction \citep{ghosh96, ghosh98}.

Now let us consider two main models and their applicability to  
high mass X-ray binary GRO J2058+42. These are namely:
Keplerian frequency and Beat frequency model \citep{klis00}.
The Keplerian frequency model, where QPOs are  produced due to 
inhomogeneities at the inner edge of the Keplerian disk,
that modulate the  X-ray flux at Keplerian frequency, 
expressed as $\nu_{\mbox{QPO}}=\nu_{k}$ \citep{klis87}. Whereas  
in the case of Beat frequency model, the accretion flow on to 
the neutron star is modulated at the Beat frequency between Keplerian 
frequency at the inner edge of the accretion disk and the neutron star 
spin frequency, $\nu_{\mbox{QPO}}=\nu_{k} - \nu_{s}$ \citep{alpar85}.
In the  case of GRO~J2058+42, the detected QPO frequency 
$ \nu_{QPO}= 9.0 \times 10^{-2}$ Hz is much   higher than the spin 
frequency  of its neutron star, $5.15 \times 10^{-3} $~Hz. It is, therefore, 
not possible to differentiate  between  a Beat frequency and a 
Keplerian frequency model in the case of GRO~J2058+42.

We can obtain radius of the inner edge of the accretion disk $r_{\mbox{QPO}}$ using expression for 
Keplerian orbital motion,
\begin{equation}
	r_{\mbox{QPO}}= \left({\frac{GM}{4\pi^2}} \right)^{1/3} \nu_{k}^{-2/3} \approx 8.0 \times 10^{8} \quad \mbox{cm},
	\label{eq:rqpo}
\end{equation} 
where $M$ is $1.4 M_{\odot}$ for a neutron star and $G$ is the gravitational constant and $\nu_{k}$
its Keplerian frequency.

  Alternatively, using mass accretion rate and strength of magnetic field of neutron star 
one can also derive radius of the inner edge of the Keplerian disk, 
which is equivalent to the magnetospheric radius $r_{m}$, of the neutron star as expressed 
by the equation 6.18, page 158 of \citet{frank02}
\begin{equation}
 r_{m} \simeq  5.2~\mu_{30}^{4/7} {\dot{M}_{16}}^{-2/7} {m_{x}}^{-1/7}
 \times 10^{8} \, \mbox{cm} = 2.4 \times 10^{8} \, \mbox{cm},
	\label{eq:rmag}
\end{equation} 
where $m_{x}= M_{x}/{M_{\odot}} = 1.4$, $\mu_{30}$ is magnetic moment of the neutron star
expressed in units of $10^{30}$ G cm$^{3}$ and $\dot{M}_{16}$ is its mass 
accretion rate expressed in units of $10^{16}$ g s$^{-1}$.  These   are  derived 
from observed values of neutron star magnetic field and source luminosity respectively
as detected by AstroSat observation.

 The co-rotation radius $r_{co}$,  of an X-ray pulsar can be defined where the spin 
angular velocity of  neutron star is equal to the Keplerian angular velocity of matter. 
It can be derived by equating the Keplerian velocity to the co-rotating Keplerian velocity,
\begin{equation}
	r_{co} = 1.7 \times 10^{8} P^{2/3} \left(\frac{M}{1.4 M_\odot} \right)^{1/3} \, \mbox{cm},
	\label{eq:rco}
\end{equation} 
where $P$ is the spin period of the neutron star. Using estimated pulse period  
from AstroSat observation and 
assuming a neutron star mass of $M = 1.4 M_{\odot}$, one  can obtain the co-rotation 
radius $r_{co}$ for GRO~J2058+42 as $5.7 \times 10^{9}$ cm (Equation-\ref{eq:rco}). 
It is therefore evident that the disk radius $r_{\mbox{QPO}}$ is  about an order of 
magnitude smaller than the co-rotation radius $r_{co}$.  It suggests, 
therefore, that  formation of such 
a transient disk is possible between magnetosphere and co-rotation radius of the neutron 
star.  Estimated  values  of the radius of the inner accretion disk derived by Keplerian 
orbital motion is ${8.0\times 10^{8}}$ cm (Equation-\ref{eq:rqpo})  and using  the accretion 
torque theory, it is determined to be ${2.4 \times 10^{8}}$ cm (Equation-\ref{eq:rmag}). 
These values  are comparable considering the level of approximations involved,
and uncertainties in mass, radius and distance of the source as well as in the geometry
of the magnetic field.  
It favors however,  formation of a transient accretion disk around 
the neutron star, which could possibly explain the cause of  observed 
oscillations at 0.090 Hz  in the X-ray flux  due to either 
Beat frequency or Keplerian frequency modulations as discussed above. Such properties 
are quite  recurrent in many cases of Be-binary pulsars as mentioned above and 
also observed in this case, for the first time by LAXPC on-board AstroSat.
The neutron star  magnetospheric radius established 
for some cases \citep{nespoli11, devasia11} using measured strength of
magnetic field, source distance and its luminosity, assuming the canonical mass and radius of a 
typical neutron star (Equation-\ref{eq:rmag}) was found to be very close to  inner radius of the 
accretion disk, as determined from its observed  QPO frequency (Equation-\ref{eq:rqpo}) for some sources. Hence,  for such 
cases, it was possible to establish consistency  with  spectroscopic measured value of their magnetic field strength.  
The same appears to be true for GRO J2058+42 under uncertainty involved in measured parameters and standard assumptions.

Formation of a transient disk may supply  necessary spin-up torque 
to the neutron star, if the disk rotates in the same direction as the pulsar spin.  
The expected torque ($N_{\mbox{char}}$) on to the neutron star due to  transfer 
of angular momentum of the accreted mass from  such a transient disk can be calculated 
using the expression below, assuming,  transfer of angular momentum  of all accreted mass from  
such a disk  with a radius $r_{QPO}$ to its neutron star \citep{zand98, mukerjee01}
\begin{equation}
	N_{\mbox{char}} = \eta {\dot{M}}(GM{r_{QPO}})^{1/2},
	\label{nchar}
\end{equation}
where $\dot{M}$ is the mass accretion rate and $\eta$ is the duty cycle
for the applied torque. 

If we assume that all the potential energy of the accreted mass liberated during the  
outburst is transformed into radiation then  the  luminosity  can be expressed as,
\begin{equation}
 L_{37} = 1.33 {\dot{M}_{17}} (M/M_{\odot}){R_{6}}^{-1} \quad \mbox{erg s}^{-1},
 \label{l37}
\end{equation}
where $L_{37}$ is the luminosity in units of $10^{37}$ erg s$^{-1}$ and $R_6$ is the
radius in units of $10^6$ cm.
The mass accretion rate can be calculated from the observed luminosity, which is
derived by measurements of flux from  spectral model for AstroSat spectrum (Table~2), 
and known distance of the star. The estimated value of $\dot{M}$ during the  
AstroSat observation was $2.3 \times 10^{17}$ g s$^{-1}$. Using these parameters, 
the value of $N_{char}$ is found to be $\eta~(9.0 \times 10^{34})$ g~cm$^{2}$~s$^{-2}$.

The observed torque $N_{0}$ of the pulsar can be expressed  in terms of the moment 
of inertia, $I= 10^{45}$ g cm$^{2}$ (for a typical neutron star, having mass as 
1.4${M_{\odot}}$ and radius  of 10 km) and  $\dot{\nu}$ the  observed rate of change 
of frequency of the neutron star as 
\begin{equation}
 N_{0} =  2{\pi}I{\dot{\nu}}.
 \label{n0}
\end{equation}
Now, if we consider the value of $\eta\approx 1.0$, representing that the transient 
accretion disk was present and torque was applied for almost the whole 
duration of observation. Then, by equating the  above two torques (Eqs.~\ref{nchar}
and \ref{n0}), 
one can estimate an average spin-up rate of  the pulsar as 
$\dot{\nu} = 1.43 \times 10^{-11}$~Hz~s$^{-1}$ or 
$\dot{P} = -5.39 \times 10^{-7}$ s~s$^{-1}$.
This is comparable to the observed value of $\dot\nu$ during AstroSat observation.

Considering the value of $\dot\nu=1.65\times10^{-11}$ Hz s$^{-1}$ obtained
during the AstroSat observation and assuming that the spin-up rate is proportional
to the luminosity during the recent outburst we could integrate $\dot\nu$ over the
entire outburst to calculate the net change in spin frequency or the period.
Taking into account the observed period of 194.22 s during the beginning of
AstroSat observation
we could calculate that the period decreased from about 195.7 s before outburst
($\mbox{MJD}=58550$) to 193.4 s 
after this latest long outburst of 2019 ($\mbox{MJD}=58624$). This change in period is comparable to
the measured period change  from 195.6 to 193.5  by Fermi-GBM during 
this latest outburst.  This is also comparable to the period change during earlier outburst observed by BATSE in 1995.
Thus, the formation and presence of such a transient accretion disks around neutron 
star during the source outburst,  could  cause  significant change in the pulsar 
period in short duration of 46 days.

\subsection{Pulse Phase-averaged Spectrum}

The  X-ray spectrum  of the source  was first derived by 
RXTE during the earlier outburst of 1996.  
The RXTE PCA  and HEXTE data were used
covering  2.7--25 keV and  11--50 keV respectively \citep{wilson05}. 
The spectral data were fitted with an absorbed thermal bremsstrahlung 
model (Phabs*bremss) adequately well,  particularly at  higher 
intensities during its outburst. The source flux dropped sharply above
20 keV in most cases, therefore,  other models such as a power 
law with a high-energy cutoff, were not used with RXTE data. 
It was found that the absorption term, 
$N_{H}$  was nearly  constant with its best-fit values in the 
range (4.6--5.4)$\times 10^{22}$ cm$^{-2}$. The temperature, kT, was found 
to increase with  the source intensity, with best-fit values varying
from $10.3 \pm 0.5$ keV at $2.9\times 10^{-11}$ ergs cm$^{-2}$ 
s$^{-1}$ to $22.2\pm 0.4$ keV at  $2.6\times 10^{-10}$ ergs cm$^{-2}$ 
s$^{-1}$ \citep{wilson05}. There was, however, no report
on any cyclotron absorption feature from RXTE observations. This was missed
very likely due to observations at much lower intensity 
of the source during low intensity outbursts, which was an order of magnitude lower
than the latest outburst.

AstroSat data enabled us to derive X-ray spectrum in 0.8--70 keV energy band as shown in Figure~7. 
Spectrum was fitted reasonably well using two models. The first model was defined as
an absorbed Fermi-Dirac cutoff model along with a black-body, a Fe-emission line  
and  3 Gaussian absorption lines were introduced to model cyclotron  scattering features and
its two higher harmonics  as observed in the spectrum. The CompTT  model was used as the 
second model in combination with a Fe-emission line and 3 Gaussian absorptions lines as defined in the first model. 
The CompTT-model is  generally used for  neutron star based low mass X-ray binaries such as Z-type  and atoll sources
with relatively  lower  magnetic field  ($<10^{12}$ G) of the  neutron star \citep{ferinelli08}.  
However, the model could  also successfully  define  spectrum of some of the pulsars in Be-binaries  for
example Cep X-4 \citep{jaiswal15}, 4U 1907+09 \citep{varun19} and GRO J2058+42 \citep{mol19}.
The derived parameters from the two models are tabulated in Table~2.
The first model  estimated a  black-body temperature of $0.83 \pm 0.04$ keV and detected  presence of  
cyclotron resonance scattering feature and its harmonics. The Comptonization model on the other hand
enabled us to determine input photon  Wien-temperature  of $0.52 \pm 0.02$ keV, the plasma temperature of $8.22 \pm 0.10$ keV and
plasma optical depth  of $5.21 \pm 0.12$ for the phase-averaged spectra.  The
Wien-temperature, as per CompTT-model,  originates away from the neutron star surface and  closer to the inner 
accretion disk i.e., at the outer transition layer, hence Wien-temperature is always found to be relatively lower than the neutron star 
black-body temperature  as it originates close to inner transition layer i.e., near the surface of the neutron 
star \citep{ferinelli08}. As per the CompTT model bulk Comptonization occurs in the innermost part of the 
transition-layer region, while thermal Comptonization is dominant in the outer transition layer and presumably 
within some extended region located above the accretion disk. Similar deviations were also observed in the case
 of Cep X-4 fitted with CompTT-model and  black-body combined with  FDCUT-model \citep[Table 1 of][]{jaiswal15}. 
However,  the  centroid  energy of cyclotron absorption feature and its detected harmonics are found
to be consistent within errors for the two models (Table~2).

Some of the accretion powered X-ray pulsars showed additional features in emission 
between 10--20 keV energy band and more rarely in absorption between 8--10 keV  in their 
respective residuals when fitted with a variety of continuum models \citep{coburn02}.  
\citet{coburn02} have argued that such features may be caused by inadequacies
of continuum model rather than cyclotron resonance features. 
For example, it was  observed that a single emission line at around
14 keV can fit two features around 10 and 20 keV for Vela X-1 \citep{kreykenbohm02}, Her X-1 \citep{coburn01},  
Cep X-4 \citep{mcbride07} and  an absorption line between 8--10 keV can fit the features
for 4U 1907+09, 4U 1538-52, 4U 0352+309 as in figure~6 of \citet{coburn02}.
In the case of GRO J2058+42, an introduction of a single Gaussian emission line in this 
range of energy  could not appropriately fit the spectrum.  Referring to the results
described in Section 3.3 using  ratios of spectral counts with respect to Crab spectrum  
derived for 4 different phases  of the pulsar (Figure~6), clearly indicates  presence of  
prominent depressions around 10 keV and 20 keV for phase-1 in particular, and  its 
presence at other phases as well. Therefore, it confirms the presence of such absorption 
features in the spectral data associated with a physical origin and not due to 
any discrepancies of the continuum model as discussed above.  Additionally, it also 
excludes the  possibility of any uncertainty in the response matrix of the detector 
as the response matrix was not used to deconvolve the spectrum to calculate these ratios. 
Relative significance of these absorption features were subsequently estimated after 
modeling the data  and results are shown in Table~2 for the phase averaged case and in 
Table~3 for 4 different  pulse phases.  These observations strongly favor  presence 
and detection of  these absorption features in their respective spectra of the pulsar.

The Be-binary  pulsar 4U 0115+63  showed interesting features when its measured  
pulsed fraction was plotted against energy starting from 5 keV onwards. It showed 
gradual increase in pulsed fraction with energy along with a sharp localized decrease 
only  around 22 keV.  This localized decrease was attributed to cyclotron resonant 
scattering at the second Landau level.  No such localized  decrease was observed
for its higher harmonics likely due to low signal to noise ratio. 
Such decrease was also  not  observed  corresponding to  its fundamental line, 
despite its high count rate, as it could be due  to competing effects such as photon 
spawning and cyclotron emission \citep{ferrigno09}.  The pulse fractions  measured 
at different  energies between 3--80 keV  from  AstroSat observation of GRO J2058+42 
(Figure~\ref{rmspf}) did not show  any localized decrease in its pulse-fraction  
corresponding to  fundamental ($\sim10$ keV) or even for  its observed higher harmonics 
($\sim20$ keV, $\sim38$ keV).  These could possibly be due to similar reasons  
mentioned as above  for 4U 0115+63 for fundamental and higher harmonics or  the 
resultant decrease in pulse fraction is  small and it could not be detected within the 
error limits.  This is very similar to another Be-binary V 0332+53,
where no  such  localized change in pulsed-fraction was observed corresponding to 
its cyclotron line energies, despite their prominent optical depths \citep{tsygan10}. 
The pulse fraction for GRO J2058+42, however, showed a gradual increase  with energy  and then became nearly 
constant at higher energies above 20 keV as seen in Figure~4.

\subsection{Pulse Phase-resolved Spectra}

Cyclotron  resonant scattering features  are produced due to interaction of photons with 
electrons quantized in the Landau states 
formed in the accretion column in the presence of a strong magnetic field  of a neutron star in  
accreting X-ray binaries \citep{schw17}.  It can also be produced by
reflection of X-rays from the atmosphere of the neutron star \citep{pout13}. 
These interactions results in absorption features in their spectrum  
at a particular energy and produce  complex shape of the  absorption features due to complex
scattering cross-sections.  The line energies and their shapes  change depending 
on the environment of the line-forming region typically located close to the neutron star,  
strength of magnetic field  of the neutron star, nature of the accretion 
column,  source luminosity and  pulse-phase of accreting X-ray pulsars \citep{nishi03, nishi05, nishi13}.

The pulse phase-resolved spectra corresponding to 4 different phases of the accreting  
pulsar GRO J2058+42 are presented in Fig.~\ref{cspec}. 
AstroSat detected a cyclotron absorption feature and its two harmonics   
in its pulse  phase-resolved  spectra, which were identified in the phased-averaged spectrum.   
Spectral parameters derived from the fitted  CompTT-model for 4 different pulse phases are 
given in Table~3.  It is noticed that the  relative strengths and shape of 
these absorption features change with pulse-phase as shown in  Figure 9 for clarity. 
The line energies are found to vary within the range of (9.7--14.4) keV, (19.3--23.8) keV 
and (37.3--43.1) keV  respectively for the observed cyclotron scattering  features and its higher harmonics. 
Their respective detection significance along with false alarm probability (FAP) with respect 
to  pulse-phases are also given in Table~3. 
We noticed a higher value of centroid energy of the fundamental  cyclotron line energy at 13.17 for phase-2,  
but it was found to be consistent with  phase-3 and Phase-4 within its associated larger uncertainty. 
Overall pulse phase-resolved  spectra  of GRO J2058+42  showed  consistency  with respect to  cyclotron line energies and
continuum except for the phase-1 (Table~3).

The phase averaged spectrum (Figure~7), fitted using CompTT-model for GRO J2058+52 determined 
the cyclotron  fundamental line energy along with the harmonics  from AstroSat observation.  
Interestingly, the observed  ratios between its fundamental and the two harmonics   
are found  to cover a range of $1.93\pm0.06$ and $3.6 \pm 0.12$  respectively with a 1-$\sigma$ error limit.  
This indicates non-harmonic ratio,  observed for  higher harmonics  of cyclotron  
resonance scattering feature with respect to its fundamental line energy.
However, looking at the results from pulse phase resolved spectra in Table~3, it is clear that
the position of the spectral features as well as their strengths showed variation with
phase but when compared with phase averaged spectrum it may give misleading results. For example,
the mean position of the first feature is different in Phase-1 as compared to other phases
and its strength is highest in Phase-1. Thus the phase averaged spectrum is likely
to be biased towards the lower value in Phase-1. On the other hand, the third feature around
38 keV is not seen in Phase-1, while it is significant in other phases and the phase
averaged spectrum will reflect this value. If we consider the ratios between energies of different
features in phase resolved spectra (Table~3), the ratios for the first two features are  $1.69\pm 0.16$, 
$1.82\pm0.14$ and $1.84\pm0.11$ corresponding to phase-2, phase-3 and phase-4 respectively.
Similarly, the  ratio between
the third and the first feature, ignoring Phase-1 where the third feature is not significant,
are $3.04\pm0.17$, $3.10\pm0.23$ and $3.47\pm0.20$ for the three phases respectively.
Within the error bars these ratios are consistent with those expected for harmonics. Hence,
there is no evidence for non-harmonic ratios in GRO J2058+42.
However, non-harmonic ratios in cyclotron absorption features were  observed for some of the pulsars, for example V 0332+53 
\citep{pottschmidt05, kreykenbohm05, nakajima10}, Her X-1 \citep{enoto08} and Cep X-4 \citep{vybornov17}.  
 For some cases, marginal deviation in harmonicity could be explained by employing relativistic approximation of 
photon-electron scattering \citep{meszaros92}, but for large deviations it could possibly be  explained due to influence
of non-dipole magnetic field on the line forming region when the field strength increases 
with height \citep{nishi05, nishi13}.

The variation in cyclotron features with pulse phase could  be  due to superposition of contributions 
from a number of lines formed at different heights of a line-forming region in a 
cylindrical accretion column geometry, influenced by a large gradient of its magnetic field. 
This could result in the observed variation of the centroid energy and shape
depending on  visibility of accretion column with respect to line of sight depending on system 
orientation during spin of the neutron star.  This could occur  either in the accretion column or 
in the neutron star atmosphere for an anisotropic injection of energy with an emission peak exiting in a  
particular direction \citep{nishi15}.  
In such cases,  cyclotron scattering features could be observed  only for the interval  
depending on its pulse-phase as seen for pulse phase-1 for GRO J2058+42.  
There could be other possibilities  where the accretion column  itself could partially be eclipsed
by the neutron star itself  for certain pulse-phases, such that only a portion  of
the accretion column  could be visible to the observer \citep{mush18}. 
This may cause dispersion in magnetic field strength across the visible 
portion of the accretion column and  may cause resultant change in the 
line-energy and its shape with pulse-phase.  This suggest  that the  
observed  phase dependent changes in the cyclotron line energy and  its shape 
for GRO J2058+42 during AstroSat observation, could likely be  due to  changes 
in the geometry of the line producing region with respect to line of sight  
during its spin, for a stable luminosity of the source.

We find from Table~3 that photo electric absorption and 
input soft photon temperature CompTT(T0) for all the 4 phases were almost constant within 
90\% error limit. The plasma temperature CompTT(kT) for phase-1 
was, however, found to be relatively higher by $\sim1.5$ keV as compared to 
an average of the remaining 3 phases at  8.5 keV.  The plasma optical depth for 
phase-2 was  found to be higher at  6.17 compared to its average value of 4.8 for the other 3 phases.
These measurements indicated that for the phase-1, the difference in  plasma temperature  and 
possibly local change in configuration of its magnetic field  could be responsible for the observed  
change in its  continuum   and line parameters as  compared to the other 3 phases. 
The  observed ratio of the spectral counts with respect to Crab for phase-1 also clearly showed that there 
are significant changes in  the spectrum relative to other phases (Figure~\ref{crabr}). 
Therefore, the evident differences in its continuum and line parameters indicate that for this
narrow phase-1,  emission probably comes mainly from a different region and hence emission components are 
different with respect to the rest of the three phases.  
For example, the radiation could originate from one column 
instead of another, from its visible higher column height as opposed to near the stellar surface.  
The stellar surface area associated with a relatively hotter plasma accumulation  may also be 
suggestive of a scenario where a magnetically more intense area produces a hotter region contributing 
a substantial variation in the spectral shape due to the interaction of accelerated high energy particles 
which could  up-scatter soft photons giving rise to the Comptonization spectrum with change 
in its spectral shape. 
The recent work of \citep{nishi19} could  explain more favorably the observed change 
in spectral continuum as well as decrease in the centroid energy of the cyclotron lines in general  and 
in particular that of the phase-1. The velocity of bulk motion of in-falling plasma in the cyclotron line forming regions
plays a dominant role.  Line forming  regions are located  near the walls of the cylindrical 
accretion column  whereas spectral continuum are formed above and around  the accumulated mound close to neutron star surface. 
Therefore,  relative location of these  two regions where continuum and cyclotron-lines are formed with respect 
to a line of sight of the observer, one could see the observed change in the spectral continuum as well as a decrease 
in centroid energy of the cyclotron-line during certain  pulse-phase of the pulsar  depending on optimal conditions such 
as velocity of the bulk motion of the in-falling matter, effect of gravitational bending of the emitted radiation and  
suitable variation of the local magnetic field \citep{nishi19}. 
The fundamental centroid energy of the cyclotron absorption 
feature showed overall variation of $\sim$30\%  with pulse-phase measured for GRO J2058+42. 
There are several other sources, namely Cen X-3 \citep{burderi00, suchy08}, 
Vela X-1 \citep{barbera03, kreykenbohm02}, 4U 0115+63 \citep{heindl04},
GX 301--2 \citep{kreykenbohm04, suchy12}, and Her X-1  \citep{voges82, soong90, klochkov08}, 
which showed  variation in their fundamental energy of about 10--30\% over 
the pulse-phase.  The  cyclotron absorption features of GRO J2058+42 clearly showed comparable variation in  cyclotron 
line-energy and  shape  with  variation in its pulse-phase.

\subsection{Determination of the Strength of Magnetic Field of the Pulsar}
The strength of the surface magnetic field B of a neutron star  can be obtained using
\begin{equation}
	E_{c} \simeq  11.6  n  \bigg({\frac{1}{1+z}} \bigg)   \bigg({\frac{B}{10^{12}G}} \bigg)\;  \mbox{keV}.
	\label{ecycl}
\end{equation}
We assume that the observed  cyclotron absorption  feature at $E_{c}$ is associated with its fundamental 
line ($n = 1$), and considering a gravitational redshift of $z = 0.3$ at the surface of a 
typical neutron star having  a mass of $1.4 M_{\odot}$ and a radius of 10 km. 
From the measurements of the 
centroid energy at $10.81^{+0.48}_{-0.49}$ keV, for the phase averaged case, we could  thus establish 
the strength  of the magnetic field of the pulsar as 
$(1.21^{+0.05}_{-0.06}) \times 10^{12}$ G. Considering the variation in energy of
first cyclotron line with  pulse phase we can get a range of (9.7--14.4) keV after including
the error bars. This would translate to a variation in the magnetic field strength 
of (1.1--1.6)$\times 10^{12}$ G over the pulse phase of the pulsar.
The strength of  the magnetic field derived for GRO~J2058+42 from its spectrum is 
comparable to the magnitude of those measured for other X-ray pulsars, 
such as KS 1947+300 (12.2 keV) \citep{furst14}, Swift J1626.3--5156 (10 keV) \citep{decesar13} and 
4U 0115+63 (12 keV) \citep{jun12, ferrigno09}  where  respective energy values of their fundamental 
cyclotron absorption features are shown within parenthesis.

\section{Conclusion}

AstroSat observed the Be-binary pulsar GRO J2058+42 on April 10, 2019  
during its latest long outburst  between March--May 2019, using 
both LAXPC and SXT instruments for a total exposure of 57 ks. The source intensity 
during this observation  had declined  to  170 mCrab, about 66\% of its 
peak intensity of 256 mCrab.  AstroSat observation 
showed a  clear detection of strong pulsation from the source.  The 
spin-period of the pulsar was determined as $194.2201\pm0.0016$ s and its 
average spin-up rate was $\dot{\nu} = (1.65\pm 0.06) \times 10^{-11}$~Hz~s$^{-1}$
corresponding to MJD 58583.10868148. 
Pulse-profiles derived  at different energy bands covering 3--80 keV, 
showed pronounced and  multiple pulse structure, which showed a variation 
in shape and pulse-fraction with energy.  The RMS pulse fraction 
varied  from about 20\% to 30\%  between  3--20 keV, beyond which it was 
approximately constant.
Pulse-profiles derived from recent AstroSat observations were found  to be
similar in shape to those reported from RXTE observations, at a relatively lower
source intensity by an order of magnitude. Thus, the source did not show drastic change in
its pulse shape and pulse-fraction within the observed  variation of 
source luminosity.

A  broad QPO feature corresponding to a frequency of 0.090 Hz  
was detected for the first time from  AstroSat observations.  
This  provided an evidence for the formation of a transient accretion disk 
around the neutron star. The QPO, therefore, offered us a tool to  probe  
inner region of the accretion disk and  to quantify the transfer 
of angular momentum through such accretion disk. This enabled us to 
estimate the torque applied to the neutron star by mass transfer from 
a transient accretion disk.  This  resulted in the  prediction and  determination of  
observed change in spin-period of the pulsar by 2.3 seconds during the recent  
46 days outburst. Similar change in  spin-period was  observed  
by BATSE during the earlier outburst in 1995 and also by  the Fermi-GBM during this
latest outburst.  This therefore, favors a scenario where pulsar could be effectively 
spun-up by such a magnitude, in a short span of 46 days.
 
The  centroid  energy of cyclotron line and its harmonics were found to be
consistent within errors for the two models used for spectral fits. The line energies are found to vary within
the range of  (9.7--14.4) keV, (19.3--23.8) keV and (37.8--43.1) keV  respectively for the observed 
3 absorption features. 
The detection of cyclotron line  lead 
to determination of strength of its strong magnetic field of the neutron star. 
Therefore, using AstroSat observation, we could establish the strength  of 
magnetic field of the pulsar as  (1.1--1.6)$\times 10^{12}$ G.
Further observations  are required to study spectral variation with source 
luminosity, variation of magnetic field strength during its 
long outburst duration and with time, to probe the nature of its accretion column  and
geometry where cyclotron lines are produced. 

\acknowledgments

We gratefully acknowledge all the support received from the Indian Space 
Research Organization (ISRO) for successful realization of AstroSat mission 
from the initial  phase of instrument building, tests and qualifications to software
developments and  mission operations. We also acknowledge the support 
received from the LAXPC Payload Operation Center (POC), TIFR, Mumbai 
for the release of verified data, calibration  data products and 
pipeline processing tools. This work has  also utilized the data from 
the Soft X-ray Telescope (SXT) and hence thankfully  acknowledge 
SXT POC at TIFR  for releasing the data  through the ISSDC data archive  center 
and providing the necessary software tools. We
acknowledge software engineers 
Sanket Kotak,  Ashutosh Bajpai and Harshal Pawar, for their vital 
services in  software development activities and timely completion of the 
full SXT pipeline  processing chain along with relevant documentation. 
We also acknowledge the support of the Department of Atomic Energy, Government of India, under
project~no.~12-R\&D-TFR-5.02-0200.  This research has  also made use 
of data obtained  from Swift-BAT and RXTE through the High Energy Astrophysics 
Science Archive Research Center On-line Services, provided by the NASA/Goddard 
Space Flight Center, we acknowledge their vital support. We also thank Fermi-GBM 
team of NASA/MSFC for sharing monitoring data of the pulsar period measurements 
of this source during its outburst. We also acknowledge  
generous support of  NASA's HEASARC  for offering all the  useful software  
and tools  for analysis of  Astronomical data. We thank an anonymous 
Referee for critical comments, which have improved the manuscript significantly.

\facility{Astrosat}


\begin{thebibliography}{}

\bibitem[Alpar \& Shaham(1985)]{alpar85}
Alpar, M. A. \& Shaham, J. 1985, Nature, 316, 239
\bibitem[Angelini et al.(1989)]{angelini89}
Angelini L., Stella L. \& Parmar A. N., 1989, ApJ, 346, 906
\bibitem[Antia et al.(2017)]{antia17}
Antia, H. M. et al.~2017 ApJS, 231, 10
\bibitem[Barthelmy et al.(2019)]{barthelmy19}
Barthelmy, S. D., et al.~2019, GCN-23985, March 22, 2019
\bibitem[Barret \& Vaughan (2012)]{barret12}
Barret, D. \& Vaughan, S. 2012, ApJ, 746, 131B.
\bibitem[Bellm et al.(2014)]{bellm14}
Bellm, E.C. et al. 2014, ApJ, 792, 108.
\bibitem[Bhalerao et al.(2015)]{bhaler15}
Bhalerao, V. et al.~2015, MNRAS, 447, 2274-2281
\bibitem[Bildsten et al.(1997)]{bildsten97}
Bildsten, L. et al.~1997, ApJS, 113, 367.
\bibitem[Bodaghee et al.(2016)]{bodag16}
Bodaghee, A. et al. 2016, ApJ, 823, 146.
\bibitem[Brejnholt et al.(2012)]{brejnholt12}
Brejnholt, N. F. et al. 2012, SPIE, vol 8843.
\bibitem[Coburn W.(2001)]{coburn01}
Coburn, W. ~2001, PhD. Thesis.
\bibitem[Coburn et al.(2002)]{coburn02}
Coburn, W. et al.~2002, ApJ, 580, 394.
\bibitem[Burderi et al.(2000)]{burderi00}
Burderi, L. et al. 2000, ApJ, 530, 429.
\bibitem[DeCesar et al.(2013)]{decesar13}
DeCesar, M. E., et al.~2013 APJ, 762, 61 
\bibitem[Devasia et al.(2011)]{devasia11}
Devasia, J., et al.~2011 MNRAS, 414, 1023 
\bibitem[Dugair et al.(2013)]{dugair13}
Dugair, M. R., et al.~2013 MNRAS, 434, 2458 
\bibitem[Enoto et al.(2008)]{enoto08}
Enoto T. et al.~2008, PASJ, 60, S57
\bibitem[Ferinelli et al.(2008)]{ferinelli08}
Ferinelli R. et al.~2008, ApJ, 680, 602
\bibitem[Ferrigno et al.(2009)]{ferrigno09}
Ferrigno C. et al.~2009, A\&A, 498, 852
\bibitem[Finger et al.(1996)]{finger96}
Finger, M. H., Wilson, R. B., \& Harmon, B. A. 1996, ApJ, 459, 288
\bibitem[Finger et al.(1998)]{finger98}
Finger, M. H., 1998, Adv. Space Res. vol 22, No.7, 1007 
\bibitem[Frank et al.(2002)]{frank02}
Frank, J., King, A., \&  Raine, D. J. 2002  Accretion Power in Astrophysics, 
    pp. 398. ISBN 0521620538. Cambridge, UK: Cambridge University Press, February 2002
\bibitem[Furst et al.(2014)]{furst14}
Furst F. et al. 2014, ApJ, 784, L40.
\bibitem[Ghosh(1996)]{ghosh96}
Ghosh, P. 1996, ApJ, 459, p 244-248.
\bibitem[Ghosh(1998)]{ghosh98}
Ghosh, P. 1998, Adv. Space Research, vol 22, No 7, p 1017-1024.
\bibitem[Harison et al.(2013)]{harison13}
Harison, F. A. et al. 2013, ApJ, 770, 103.
\bibitem[Heindl et al.(2004)]{heindl04}
Heindl, W. A. et al. 2004, in AIP Conf. Proc.714, X-RAY TIMING 2003: Rossie and Beyond, ed. P. Kaaret, F. K. Lamb, \& J. H. Swank (Melville, NY: AIP), 323
\bibitem[Heindl et al.(1999)]{heindl99}
Heindl, W. A. et al. 1999, ApJ, 521, L49-53.
\bibitem[In't Zand et al.(1998)]{zand98}
In't Zand, J. J. M., Baykal, A., \& Strohmayer, T. E. 1998, ApJ, 496, 386.
\bibitem[Jaiswal et al.(2015)]{jaiswal15}
Jaiswal, G. K. et al. ~2015, MNRAS, 453, L21-25. 
\bibitem[Jun et al.(2012)]{jun12}
Jun, L. et al. ~2012, MNRAS, 423, 2854
\bibitem[Kennea et al.(2019)]{kennea19}
Kennea, J. A. et al.~2019, GCN-24021, March 27, 2019
\bibitem[Klochkov  et al.(2008)]{klochkov08}
Klochkov, D.  et al.,~ 2008, A\&A 482,907 
\bibitem[Kreykenbohm et al.(2002)]{kreykenbohm02}
Kreykenbohm, I.  et al.,~ 2002, A\&A 395, 129
\bibitem[Kreykenbohm et al.(2004)]{kreykenbohm04}
Kreykenbohm, I.  et al.,~ 2004, A\&A 427, 975 
\bibitem[Kreykenbohm et al.(2005)]{kreykenbohm05}
Kreykenbohm, I.  et al.,~ 2005, A\&A 433, L45 
\bibitem[Krimm et al.(2013)]{krimm13}
Krimm, H. A. et al.,~ 2013, ApJSS 209, 14
\bibitem[La Barbera et al.(2001)]{barbera01}
La Barbera, A., et al.~2001 ApJ, 553, 375 
\bibitem[La Barbera et al.(2003)]{barbera03}
La Barbera, A., et al.~2003 A\&A, 400, 993 
\bibitem[Maitra(2016)]{maitra16}
Maitra, C. ~2016, J Ap. Astr., 123, 23
\bibitem[Malacaria et al.(2019)]{malac19}
Malacaria, C., Jenke, P., Wilson-Hodge, C. A., \& Roberts, O. J.~2019, ATel~12614, March 28, 2019
\bibitem[McBride et al.(2007)]{mcbride07}
McBride V. A. et al. ~2007, A\&A, 470, 1065.
\bibitem[Meszaros, et al.(1992)]{meszaros92}
Meszaros, P.~ 1992, High-Energy Radiation from Magnetized Neutron Stars (Chicago, IL: Univ. Chicago Press)
\bibitem[Molkov et al.(2019)]{mol19}
Molkov, S., Lutovinov, A., Tsygankov, S., Mereminskiy, I., \& Mushtukov, A.~2019, ApJ, 883, L11
\bibitem[Mukerjee et al.(2000)]{mukerjee00}
Mukerjee, K., Agrawal, P. C. et al.~2000, A\&A, 353, 239
\bibitem[Mukerjee et al.(2001)]{mukerjee01}
Mukerjee, K., Agrawal, P. C. et al.~2001, ApJ, 548, 368
\bibitem[Mushtukov et al.(2018)]{mush18}
Mushtukov, A. A.~2018, MNRAS, 474, 5425
\bibitem[Nakajima(2010)]{nakajima10}
Nakajima, M.~2010, ApJ, 710, 1755
\bibitem[Nespoli \& Reig (2011)]{nespoli11}
Nespoli, E. \& Reig, P. ~2011, A\&A, 526, A7 
\bibitem[Nishimura(2003)]{nishi03}
Nishimura, O.~2003, PASJ, 55,849 
\bibitem[Nishimura(2005)]{nishi05}
Nishimura, O.~2005, PASJ, 57, 769
\bibitem[Nishimura(2013)]{nishi13}
Nishimura, O.~2013, PASJ, 65,84 
\bibitem[Nishimura(2015)]{nishi15}
Nishimura, O.~2015, ApJ, 807, 164
\bibitem[Nishimura(2019)]{nishi19}
Nishimura, O.~2019, PASJ, 71(2), 42(1-9) 
\bibitem[Okazaki et al.(2001)]{okazaki01}
Okazaki A. T. and Negueruela I. ~2001, A\&A, 377, 161
\bibitem[Okazaki et al.(2002)]{okazaki02}
Okazaki A. T., Bate M. R., Ogilvie G. I. and Pringle J. E. ~2002, MNRAS, 337, 967
\bibitem[Okazaki(2016)]{okazaki16}
Okazaki A. T. ~2016, APS Conference series, Vol 506, 1 
\bibitem[Orlandini et al.(2012)]{orlan12}
Orlandini, M. et al.~2012, ApJ, 748, 86
\bibitem[Parmar et al.(1989)]{parmar89}
Parmar, A., et al.~1989, ApJ, 338, 373
\bibitem[Paul \& Rao(1998)]{paul98}
Paul, B., \& Rao, A. R. 1998, A\&A, 337, 815
\bibitem[Paul et al.(2001)]{paul01}
Paul, B., et al.~2001, A\&A 370, 529
\bibitem[Protassov et al.(2002)]{prota02}
Protassov, R., et al.~2002, ApJ 571, 545
\bibitem[Pottschmidt et al.(2005)]{pottschmidt05}
Pottschmidt, K., et al.~2005, ApJ 634, L97 
\bibitem[Poutanen et al.(2013)]{pout13}
Poutanen, J., et al.~2013, ApJ 777, 115
\bibitem[Qui et al.(2005)]{qu05}
Qu L. J.. et al.~2005, ApJ, 629, L33-L36 
\bibitem[Reig et al.(2011)]{reig11}
Reig, P.,~2011 Ap \& SS 332, issue 1, p 1-29
\bibitem[Reig et al.(2005)]{reig05}
Reig, P., Negueruela I., Papamastorakis, G., Manousakis, A.\& Kougentakis, T., 2005 A\&A 440, 637-646
\bibitem[Schwarm et al.(2017)]{schw17}
Schwarm, F.-W. et al.~2017, A\&A 601, A99
\bibitem[Singh et al.(2016)]{singh16}
Singh, K. P., et al.~2016, Proceedings of the SPIE, Volume 9905, id. 99051E 10 pp.
\bibitem[Soong \& Swank(1989)]{soong89}
Soong, Y., \&  Swank, J. H. ~1989, Proc. 23rd ESLAB Symp. on Two Topics in X-ray Astronomy,
eds. J. Hunt, \& B. Battrick, ESASP 296..617S  
\bibitem[Soong  et al.(1990)]{soong90}
Soong, Y., et al.~1990, ApJ, 348, 641
\bibitem[Staubert et al.(2019)]{staubert19} 
Staubert, R., Trumper, J., Kendziorra, E., et al. 2019, A\&A, 622, A61
\bibitem[Suchy et al.(2008)]{suchy08}
Suchy, S., et al.~2008, ApJ, 675, 1487
\bibitem[Suchy et al.(2012)]{suchy12}
Suchy, S., et al.~2012, ApJ, 745, 124
\bibitem[Takeshima(1994)]{takeshima94}
Takeshima, T., Dotani, T., Mitsuda, K. \& Nagase, F.~1994, ApJ, 436, 871-874
\bibitem[Takeshima(1998)]{takeshima98}
Takeshima, T. 1998, Proc. IAU Symposium 188, Ed. K. Koyama, S. Kitamoto, M. Itoh,
Kluwer Academic, Dordrecht, p.~368
\bibitem[Tanaka(1986)]{tanaka86}
Tanaka, Y. 1986, in IAU Colloq. 89: Radiation Hydrodynamics in Stars and
Compact Objects, eds. D. Mihalas \& K.-H. A. Winkler, 198
\bibitem[Titarchuk(1994)]{tit94}
Titarchuk, L.~1994, ApJ 434, 570
\bibitem[Truemper et al.(1978)]{truemper78}
Truemper, L. et al.~1978, ApJ 219, L105-L110
\bibitem[Tsygankov et al.(2010)]{tsygan10}
Tsygankov, S. S.~2010, MNRAS, 401, 1628
\bibitem[van der Klis et al.(1987)]{klis87}
van der Klis, M., Stella, L., White, N., Jansen, F., \& Parmar, A. N. 1987, ApJ, 316, 411
\bibitem[van der Klis(2000)]{klis00}
van der Klis, M.~2000, Annu. Rev. Astron. Astrophys. 38:717-760
\bibitem[Varun et al.(2019)]{varun19}
Varun et al.~2019, ApJ, 880, 61 
\bibitem[Wang et al.(2014)]{wang14}
Wang, D. H., Chen, L.,  Zhang, C. M., Lei, Y. J., and Qu, J. L.,~ 2014, Astronomische Nachrichten 335, 168-177
\bibitem[Voges et al.(1982)]{voges82}
Voges, W. et al.~1982, ApJ, 263,803 
\bibitem[Vybornov et al.(2017)]{vybornov17}
Vybornov, V. et al.~2017, A\&A 601, A126
\bibitem[Wilson et al.(1998)]{wilson98}
Wilson, C. A., Finger, M. H. Harmon, B. A., Chakrabarty, D., \& Strohmayer, T.~1998, ApJ 499, 820
\bibitem[Wilson et al.(2005)]{wilson05}
Wilson, C. A. et al.~2005, ApJ 622, 1024

\end{thebibliography}
\end{document}